\documentclass[letterpaper,12pt]{article}

\usepackage{endnotes}
\usepackage{amssymb}
\usepackage{amsfonts}
\usepackage{amsmath}
\usepackage{amsthm}
\usepackage{graphicx,color}

\usepackage{comment}
\usepackage{array}

\usepackage{url}

\usepackage{setspace}

\usepackage{natbib}
\setcitestyle{numbers}
\setcitestyle{square}

\def\R{\mathbb{R}}
\def\endproof{\hfill\diamondsuit}

\def\sF{{\mathcal F}}

\def\sA{{\mathcal A}}
\def\sL{{\mathcal L}}

\def\sX{{\mathcal X}}

\def\sC{{\mathcal C}}

\def\E{\mathbb{E}}

\def\sF{\mathcal{F}}
\def\P{\mathbb{P}}

\def\N{\mathbb N}

\numberwithin{equation}{section}
\theoremstyle{plain}                
\newtheorem{theorem}{Theorem}[section]
\newtheorem{lemma}[theorem]{Lemma}
\newtheorem{proposition}[theorem]{Proposition}
\newtheorem{corollary}[theorem]{Corollary}
\theoremstyle{definition}           
\newtheorem{definition}[theorem]{Definition}

\theoremstyle{remark}               
\newtheorem{remark}{Remark}[section]

\usepackage[margin=1.25in]{geometry}

\usepackage{ulem} 

\begin{document}

\begin{center}
\large{\bf Existence of an equilibrium with limited stock market participation and power utilities}\footnote{We would like to thank G. Chabakauri, M. Larsson,  J. Ma, I. Tice, A. Shadi Tahvildar-Zadehand, and J. Zhang for constructive comments.  Kasper Larsen is corresponding author and has contract information: Email: \url{KL756@math.rutgers.edu} and mailing address: Department of Mathematics, Rutgers University, Hill Center 330 - Busch Campus, 110 Frelinghuysen Road, Piscataway, NJ 08854-8019, USA.}

\ \\

{\large \bf Paolo Guasoni}\\
Universit\`a di Bologna,\\
Dublin City University

\ \\

{\large \bf Kasper Larsen}\\
Rutgers University

\ \\ 

{\large \bf Giovanni Leoni}\\
Carnegie Mellon University
\ \\

\end{center}
\begin{center}
\ \\

{\normalsize \today }
\end{center}
\vspace{.5cm}

\begin{verse} {\sc Abstract}:  
For constants $\gamma \in (0,1)$ and $A\in (1,\infty)$, we prove existence and uniqueness of a solution to the singular and path-dependent Riccati-type ODE
 \begin{align*}
\begin{cases}
h'(y)  = \frac{1+\gamma}{y}\big( \gamma - h(y)\big)+h(y)\frac{\gamma  + \big((A-\gamma)e^{\int_y^1 \frac{1-h(q)}{1-q}dq}-A\big)h(y)}{1-y},\quad y\in(0,1), \\
h(0) = \gamma, \quad h(1) = 1.
\end{cases}
\end{align*}
As an application, we use the ODE solution to prove existence of a Radner equilibrium with homogenous power-utility investors in the limited participation model from Basak and Cuoco (1998).
\end{verse}

\vspace{0.25cm}


\newpage

\noindent Declaration of interest: The research of G. Leoni was partially supported by the National Science Foundation under Grant No. DMS 2108784. The research of P. Guasoni was partially supported by SFI (16/IA/4443). \ \\

\noindent Declaration of generative AI in scientific writing: No AI nor AI-assisted tools have been used.\ \\

\noindent Data availability statement: We do not analyze nor generate any datasets.

\newpage

\section{Introduction}
There is an extensive literature on nonlinear second-order differential
equations of the type%
\begin{equation}
\big(\varphi(y)G(w^{\prime})w^{\prime}\big)^{\prime}+\varphi(y)F(y,w,w^{\prime
})=0,\quad y\in (0,1).\label{second order}%
\end{equation}
When $G$ and $F(y,w,\cdot)$ depend only on $|w^{\prime}|$ and $\varphi
(y)=y^{n-1}$ with $y:=|x|$, equation (\ref{second order}) becomes the radial version of the
differential equation
\begin{equation}
\operatorname{div}\big(G(|Dw|)Dw\big)+F(|x|,w,|Dw|)=0,\label{pde}
\end{equation}
see, for example, \cite{ni-serrin86}, \cite{pucci-serrin91}, and \cite{pucci-serrin-book07}.
In this case, solutions to (\ref{second order}), or related differential
inequalities, are often used to prove comparison and maximum principles for the
corresponding differential equation (see \cite{pucci-serrin-book07}). Important examples are $G(\xi)=1$ and $G(\xi)=|\xi|^{p-2}$, $p>1$, which yield the
Laplace equation%
\begin{align}\label{Laplace}
\Delta u+F(|x|,u,|Du|)=0,
\end{align}
and the $p$-Laplace equation
\begin{align}\label{pLaplace}
\operatorname{div}(|Du|^{p-2}Du)+F(|x|,u,|Du|)=0,
\end{align}
respectively. In \eqref{Laplace} and \eqref{pLaplace},  we use $u(x):=w(|x|)$ and $y:=|x|$.

The differential equation (\ref{second order}) also arises in the study of singular
self-similar solutions to the porous media equation with absorption
\[
\frac{\partial u}{\partial t}=\Delta(u^{m})-u^{p},
\]
where one looks for solutions of the type%
\[
u(x,t)=t^{-a}w(t^{-b}|x|),
\]
see, for example, \cite{brezis-peletier-termam86}, \cite{iagar-laurencot13}, and \cite{leoni96}.

In this paper, we are interested in a special form of (\ref{second order}),
which arises from a system of coupled stochastic control problems in financial economics (discussed below and detailed in Section \ref{sec:application}).  For $\gamma \in (0,1)$ and $A\in (1,\infty)$, we consider the second order nonlinear differential equation
\begin{align}\label{second order2}
\big(\varphi(y)w^{\prime}\big)^{\prime}-\gamma(1+\gamma)y^\gamma(1-y)^{1-\gamma}-\varphi(y)\big(  (1-y)e^{w}-A\big)  (w^{\prime})^{2}=0,
\end{align}
for $y\in(0,1)$, where%
\[
\varphi(y):=y^{1+\gamma}(1-y)^{\gamma},\quad y \in (0,1).
\]
To highlight the singularities at $y=0$ and $y=1$, we write the differential equation \eqref{second order2} as
\begin{equation}
w^{\prime\prime}(y)=\frac{\gamma(1+\gamma)}{y(1-y)}+\frac{(1+\gamma
)(2y-1)}{y(1-y)}w^{\prime}(y)+\big((1-y)e^{w(y)}-A\big)  w^{\prime
}(y)^{2}.\label{ode 2}
\end{equation}
For our application in Section \ref{sec:application} below, we are interested in
a singular solution $w\in \sC^{2}([0,1))$, which satisfies the boundary condition
\begin{equation}
\lim_{y\uparrow1}(1-y)w^{\prime}(y)=1.\label{limit}%
\end{equation}
We will see that a necessary and sufficient condition for the local existence
of a smooth solution for $y$ near $0$ is that the boundary condition
\begin{equation}
w^{\prime}(0)=\gamma\label{bv1}
\end{equation}
holds. The main challenge of the paper is to determine the range of initial values
\begin{equation}
w(0)=w_{0}\label{bv2}%
\end{equation}
for which equation (\ref{ode 2}) admits a global solution $w(y)$ for $y\in (0,1)$ and construct a unique solution that satisfies (\ref{limit}). Following
\cite{brezis-peletier-termam86} and \cite{leoni96}, we use a shooting
argument. To be precise, we show that for all
\[
0\leq w_{0}<A-\gamma,
\]
the differential equation (\ref{ode 2}) has a unique solution satisfying the boundary conditions
(\ref{bv1}), (\ref{bv2}), and%
\[
\lim_{y\uparrow1}(1-y)w^{\prime}(y)=\frac{\gamma}{A}\in(0,1).
\]
Then, we show that for all $w_{0}$ sufficiently large, the solutions to the
Cauchy problem (\ref{ode 2}), (\ref{bv1}), and (\ref{bv2}) explode for $y<1$. Finally, we show that the initial value $w_{0}$ for which (\ref{limit}) holds is
given by the supremum of the set of initial values $w_{0}>0\,$\ for which
there is a smooth solution to (\ref{ode 2}), (\ref{bv1}), and (\ref{bv2})
satisfying
\[
\lim_{y\uparrow1}(1-y)w^{\prime}(y)\leq\gamma.
\]

The main challenges with respect to previous work are that: (i) the ODE 
(\ref{ode 2}) is singular at both $y=0$ and $y=1$,  (ii) the ODE (\ref{ode 2}) is not
variational, so standard techniques (see, e.g., \cite{mawhin-book87}) cannot be
applied, (iii) the ODE (\ref{ode 2}) is of Riccati type due to the
presence of the term $(w^{\prime})^{2}$, (iv) there is exponential growth in
$w$, and (v) we are looking for singular solutions (see, e.g.,
\cite{garcia-manasevich-yarur-98} and \cite{ni-serrin86}).

To state our main result, we rewrite (\ref{ode 2}) in terms of 
\[
h(y):=(1-y)w^{\prime}(y),\quad y\in(0,1).
\]
Then, the ODE  (\ref{ode 2}) becomes 
 \begin{align}\label{KEY_ODE}
\begin{cases}
h'(y)  = \frac{1+\gamma}{y}\big( \gamma - h(y)\big)+ h(y)\frac{\gamma  + \big((A-\gamma)e^{\int_y^1 \frac{1-h(q)}{1-q}dq}-A\big)h(y)}{1-y},\quad y\in(0,1), \\
h(0) = \gamma, \quad h(1) = 1.
\end{cases}
\end{align}
Our main mathematical result is:
\begin{theorem}\label{theorem_main_h}For $\gamma \in (0,1)$ and $A\in (1,\infty)$, there exists a unique solution $h \in \sC^1([0,1])$ of \eqref{KEY_ODE} satisfying $\gamma \le h \le 1$.
\end{theorem}

Our motivation for studying \eqref{KEY_ODE} comes from the limited stock-market participation model in \cite{BC1998}. We use Theorem \ref{theorem_main_h} to prove the  existence of a Radner equilibrium when both investors have identical power utility functions with common relative risk-aversion coefficient $\gamma \in (0,1)$ and common time-preference parameter $\beta>0$. Our analysis is significantly more involved than the log-log-utility model originally developed in \cite{BC1998}. Because of the log-utility assumption placed on the restricted investor, the model in \cite{BC1998} is explicitly solvable and no ODE is needed. 

There are several existing model extensions of \cite{BC1998}. \cite{Hug2012} proves existence of equilibrium bubbles (i.e., models where the equilibrium stock-price process is a strict local martingale) and shows that equilibrium uniqueness can fail in models with multiple stocks. In another extension,  \cite{Pri2013} proves the existence of an equilibrium when the unrestricted investor has a power-utility function. The existence proofs in \cite{BC1998},  \cite{Hug2012}, and \cite{Pri2013} all rely crucially on the restricted investor having a logarithmic utility function. This is because the optimal policy  for an investor with a logarithmic utility function is available in closed form.  \cite{Weston} uses coupled BSDEs to prove equilibrium existence for investors with heterogenous exponential utilities. Finally, our existence proof also puts some of the experimental numerics for homogenous investors reported for power-power utilities in \cite{Chab2015} on a solid mathematical foundation. 

We use the mathematical setting from \cite{BC1998}, which falls under the theory of incomplete Radner equilibria. Incompleteness stems from the restricted investor's inability to hold stocks. Utilities defined on the positive real line --- such as log and power --- restrict consumption processes to be nonnegative, which complicates the underlying mathematical structure. In continuous-time settings with noise generated by Brownian motions, incomplete Radner equilibria are originally discussed in \cite{CH1994} but no general existence result is available.\footnote{The counterpart of complete equilibrium models is fully developed; see, e.g.,  Chapter 4 in \cite{KS1998} and \cite{Anderson-Raimondo}.}

\section{ODE existence}

\subsection{Auxiliary ODE results}

\begin{lemma}\label{lemma_phi}Let $\gamma \in (0,1)$. Then, we have:
\begin{enumerate}
\item 
The function
\begin{align}\label{phi}
\varphi(y):=y^{1+\gamma}(1-y)^{ \gamma },\quad y\in[  0,1],
\end{align}
satisfies 
\begin{equation}
\frac{\varphi^{\prime}(y)}{\varphi(y)}=\frac{1+ \gamma }{y}-\frac{ \gamma }%
{1-y},\quad y\in(  0,1). \label{quotient phi}%
\end{equation}
\item The function
\[
\phi(y):=\frac{1}{\varphi(y)}\int_{0}^{y}\frac{\varphi(t)}{t}dt,\quad
y\in(  0,1),
\]
can be extended to $\phi\in \sC^{1}([0,1))  $ with 
\begin{align}\label{phi0}
\phi(0)=\frac{1}{1+ \gamma },\quad \phi^{\prime}(0)=\frac{\gamma }{(1+\gamma )(2+\gamma )}.
\end{align} 
\end{enumerate}

\end{lemma}

\proof 1: Follows by computing derivatives. \\
\noindent 2: By Taylor's formula centered at $0$, we have
\begin{align*}
\tfrac{\varphi(t)}{t}  &  =t^{\gamma }(1-t)^{ \gamma }\\
&=t^{\gamma }-\gamma t^{1+\gamma}+o(t^{1+\gamma }),\\
\tfrac{1}{(1-y)^{\gamma}}  &  =1+y\gamma  +o\left(y\right).
\end{align*}
Therefore, 
\begin{align*}
\phi(y)=\tfrac{1}{\varphi(y)}\int_{0}^{y}\tfrac{\varphi(t)}{t}dt  &  =\tfrac{1+y\gamma+o\left(  y\right)  }{y^{1+\gamma }}\int_{0}^{y}\big(t^{\gamma 
}-\gamma t^{1+\gamma }+o(t^{1+\gamma })\big)dt\\
&  =\tfrac{\big(1+y\gamma   +o\left(  y\right)  \big)\left(  \tfrac
{1}{1+\gamma }y^{1+\gamma }-\tfrac{\gamma }{2+\gamma }y^{2+\gamma }+o(y^{2+\gamma 
})\right) }{y^{1+\gamma }}\\
&  =\tfrac{1}{1+\gamma }+\gamma \left(  \tfrac{1}{1+\gamma }-\tfrac{1}{2+\gamma 
}\right)  y+o(y).
\end{align*}
This shows that $\phi(y)$ can be extended continuously to $y=0$ with the boundary values in \eqref{phi0}. For $y>0$,  by (\ref{quotient phi}), the previous calculation, and Taylor's formula, we have 
\begin{align*}
\phi^{\prime}(y)  &  =-\frac{\varphi^{\prime}(y)}{\varphi(y)^2}\int_{0}%
^{y}\frac{\varphi(t)}{t}dt+\frac{1}{\varphi(y)}\frac{\varphi(y)}{y}\\
&  =-\left(  \tfrac{1+ \gamma }{y}-\tfrac{\gamma }{1-y}\right)  \phi(y)+\tfrac
{1}{y}\\
&  =-\left(  \tfrac{1+\gamma }{y}-\gamma -y\gamma +o\left(  y\right)  \right)  \left(  \tfrac{1}{1+\gamma  
}+\gamma \big(  \tfrac{1}{1+\gamma}-\tfrac{1}{2+ \gamma }\big)
y+o(y)\right)  +\tfrac{1}{y}\\
&  =\tfrac{ \gamma }{(1+\gamma )(2+ \gamma )}+o(1)\\
&=\phi^{\prime}(0)+o(1).
\end{align*}
Hence, $\phi$'s extension is in $\sC^{1}([0,1))$.
$\endproof$

\begin{theorem}\label{theorem local existence}Let $a:\left[  0,1\right]
\rightarrow\mathbb{R}$ be a continuous function, $\gamma\in(0,1)$, and define 
\begin{align}\label{a0a1}
\begin{split}
a_0(y)&:= \frac{\gamma (1+\gamma )}{y},\quad  a_1(y) :=  \frac{(2 \gamma +1) y-(1+\gamma)}{y},\quad y\in (0,1].
\end{split}
\end{align}
Then, the singular Riccati ODE
\begin{align}
f^{\prime}(y) &  =a_0(y)+\tfrac{a_1(y)}{1-y}f(y)+\tfrac{a(y)}{1-y}f(y)^2,\quad y>0,\label{ode f}
\end{align}
with the boundary condition $f(0)=\gamma $, has a strictly positive local solution in $\sC^1([0,\delta])$ in the sense that 
there exist $\delta \in (0,\frac{1}{2}]$ and a function $f:[0,\delta
]\rightarrow(0,\infty)$ with $f\in\sC^1([0,\delta])$ satisfying \eqref{ode f} for $y\in (0,\delta)$.
\end{theorem}

\proof
\textbf{Step 1/6:} We eliminate the linear term in \eqref{ode f} by multiplying both sides of (\ref{ode f}) by $\varphi$ from Lemma \ref{lemma_phi} to see
\[
\big(\varphi(y)f(y)\big)^{\prime}=a_0(y)\varphi(y)+\frac
{a(y)}{1-y}\varphi(y)f(y)^2,\quad y>0.
\]
Integrating both sides and using $\varphi(0)=0$ and
$f(0)=\gamma $, yields
\[
\varphi(y)f(y)=\int_{0}^{y}\varphi(t)\left(  a_0(t)+\frac{a(t)}{1-t}f(t)^2\right)dt,\quad y>0.
\]
Hence, if a local solution $f$ of \eqref{ode f} exists, it satisfies the integral equation
\[
f(y)=\frac{1}{\varphi(y)}\int_{0}^{y}\varphi(t)\left( a_0(t)+\frac{a(t)}{1-t}f(t)^2\right)  dt,\quad y>0.
\]
The next steps construct a fixed point for this integral equation for $y\in [0,\delta]$ for some $\delta\in (0,\frac12]$.

\noindent \textbf{Step 2/6: }For $0<\delta\leq\frac{1}{2}$  to be chosen later and an arbitrary $R>0$, we
consider the closed convex set%
\begin{align}\label{sX}
\sX:=\{f\in \sC([0,\delta]):\,\,\Vert f-\gamma \Vert_{\infty}\leq R\},
\end{align}
where $\Vert f\Vert_{\infty}:=\max_{y\in\lbrack0,\delta]}|f(y)|$. For $f\in
\sX$ and $0<y\leq\delta$, we define $T$ by
\begin{align}\label{Tf}
T(f)(y):=\frac{1}{\varphi(y)}\int_{0}^{y}\varphi(t)\left(  a_0(t)+\tfrac{a(t)}{1-t}f(t)^2\right)  \,dt.
\end{align}
We claim that $\lim_{y\downarrow0}T(f)(y)$ exists and 
\begin{equation}
\lim_{y\downarrow0}T(f)(y)=\gamma. \label{limit at 0}%
\end{equation}
To see \eqref{limit at 0}, we use L'Hospital's rule, \eqref{quotient phi}, and \eqref{Tf}  to get
\begin{align*}
& \lim_{y\downarrow 0} \frac{\varphi(y)}{\varphi^{\prime}(y)}\left[  a_0(y)+\frac{a(y)}{1-y}f(y)^2\right] \\
&= \lim_{y\downarrow 0} \frac{\gamma(1+\gamma)+y\frac{a(y)}{1-y}f(y)^2}{1+\gamma -y\frac{\gamma }{1-y}}\\
&=
\gamma.
\end{align*}
We extend $T$ continuously to $y=0$ by defining $T(f)(0):= \gamma $ so that $T:\sX\rightarrow \sC([0,\delta])$.

\noindent\textbf{Step 3/6: }We claim that if $\delta>0$ is sufficiently small, then%
\begin{equation}
T(\sX)\subseteq \sX. \label{inclusion}%
\end{equation}
Consider the continuous function%
\begin{align}\label{def_psi}
\psi(y,z):=\tfrac{a(y)}{1-y}z^2,\quad  y\in [0,\tfrac{1}{2}],\quad z\in\lbrack\gamma -R,\gamma +R].
\end{align}
Because $\psi$ is continuous, there exists a constant $M>0$ such that
\begin{equation}
|\psi(y,z)|\leq M,\quad \forall (y,z)\in [0,\tfrac{1}{2}]\times\lbrack\gamma -R,\gamma +R]. \label{bound}%
\end{equation}
The constant $M$ depends on the function $a(y)$ and $R$ but $M$ does not depend on $\delta$. 

Provided $\delta \in(0, \frac12)$  satisfies
\[
\delta\leq\delta_{1}:=\frac{R(2+\gamma )}{2M2^{\gamma }},
\]
for $f\in \sX$ and $y\in [0,\delta]$, we have  the following estimate
\begin{align}\label{500}
\begin{split}
&  \frac{1}{\varphi(y)}\int_{0}^{y}\varphi(t)\left\vert \tfrac{a(t)}{1-t}f(t)^2\right\vert \,dt\\
&  =\frac{1}{\varphi(y)}\int_{0}^{y}\varphi(t)\left\vert \psi\big(t,f(t)\big)\right\vert \,dt\\
&\leq\frac{M}{y^{1+ \gamma }(1-y)^{\gamma }}\int_{0}%
^{y}t^{1+\gamma  }(1-t)^{\gamma }\,dt\\
&  \leq\frac{M2^{\gamma }}{y^{1+ \gamma }}\int_{0}^{y}t^{1+ \gamma }\,dt\\
&=\frac{M2^{\gamma }}{(2+\gamma )}y\\
&\leq\frac{R}{2}.
\end{split}
\end{align}
On the other hand, L'Hospital's rule, \eqref{phi}, and \eqref{a0a1} give
\[
\lim_{y\downarrow0}\frac{1}{\varphi(y)}\int_{0}^{y}\varphi(t)a_0(t)\,dt=\gamma .
\]
Hence, there exists $\delta_{2}>0$ such that
\begin{equation}
\forall  y\in (0,\delta_2): \left\vert \frac{1}{\varphi(y)}\int_{0}^{y}\varphi(t)a_0(t)\,dt- \gamma \right\vert \leq\frac{R}{2}. \label{501}%
\end{equation}
Taking $\delta\leq\min\{\delta_{1},\delta_{2},\tfrac12\}$, it follows from the
estimates (\ref{500}) and (\ref{501}) that%
\[
|T(f)(y)-\gamma |\leq R,\quad \forall y\in\lbrack0,\delta],\quad \forall f\in \sX.
\]
Therefore, the inclusion (\ref{inclusion}) holds.

\noindent \textbf{Step 4/6: }This step proves that the operator $T$ from step 2 is equi-continuous (even better, $T$ turns out to be an equi-Lipschitz operator). For
$0<y_{1}<y_{2}\leq\delta$ and $f\in \sX$, the formula for $a_0(y)$ in \eqref{a0a1}, \eqref{Tf}, and \eqref{def_psi} give
\begin{align}\label{502}
\begin{split}
&|T(f)(y_{2})    -T(f)(y_{1})|\\
&\leq(1+\gamma )\gamma \left\vert \frac
{1}{\varphi(y_{2})}\int_{0}^{y_{2}}\frac{\varphi(t)}{t}dt-\frac{1}%
{\varphi(y_{1})}\int_{0}^{y_{1}}\frac{\varphi(t)}{t}dt\right\vert \\
&  \quad+\left\vert \frac{1}{\varphi(y_{2})}\int_{0}^{y_{2}}\varphi
(t)\psi\big(t,f(t)\big)\,dt-\frac{1}{\varphi(y_{1})}\int_{0}^{y_{1}}\varphi
(t)\psi\big(t,f(t)\big)\,dt\right\vert \\
&  \leq(1+\gamma )\gamma \left\vert \frac{1}{\varphi(y_{2})}\int_{0}^{y_{2}%
}\frac{\varphi(t)}{t}dt-\frac{1}{\varphi(y_{1})}\int_{0}^{y_{1}}\frac
{\varphi(t)}{t}dt\right\vert \\
&  \quad+\frac{1}{\varphi(y_{2})}\int_{y_{1}}^{y_{2}}\varphi(t)|\psi
\big(t,f(t)\big)|\,dt+\left\vert \frac{1}{\varphi(y_{2})}-\frac{1}{\varphi(y_{1}%
)}\right\vert \int_{0}^{y_{1}}\varphi(t)|\psi\big(t,f(t)\big)|\,dt\\
&  =:I+II+III.
\end{split}
\end{align}
We estimate each of the three terms separately. Lemma \ref{lemma_phi} ensures $\phi\in
\sC^{1}([0,\delta])$, and so, the Mean-Value Theorem gives
\[
I\leq(1+\gamma )\gamma (y_{2}-y_{1})\Vert\phi^{\prime}\Vert_{\infty}.
\]
The bound in (\ref{bound}) and $y_{2}\leq\delta\leq\frac{1}{2}$ give
\begin{align*}
II  &  \leq M\frac{1}{y_{2}^{1+ \gamma }(1-y_2)^{\gamma }}\int_{y_{1}}%
^{y_{2}}t^{1+\gamma }(1-t)^{ \gamma }dt\\
&  \leq M\frac{2^{ \gamma }}{y_{2}^{1+ \gamma }}\int_{y_{1}}^{y_{2}}t^{1+\gamma}dt\\
&\leq M2^{\gamma }(y_{2}-y_{1}).
\end{align*}
Similar, the bound in (\ref{bound}) and $0<y_{1}<y_{2}\leq\delta\leq\frac{1}{2}$ give
\begin{align*}
III  &  \leq\frac{|\varphi(y_{1})-\varphi(y_{2})|}{\varphi(y_{2})\varphi
(y_{1})}M\int_{0}^{y_{1}}\varphi(t)\,dt\\
&\leq2^{2\gamma}\frac{|\varphi (y_{1})-\varphi(y_{2})|}{y_{2}^{1+ \gamma }y_{1}^{1+ \gamma }}M\int_{0}^{y_{1}
}t^{1+\gamma}dt\\
&  \leq2^{2\gamma}\frac{|y_{1}-y_{2}|}{y_{2}^{1+\gamma }}|\varphi^{\prime
}(z)|My_{1}\\
&\leq2^{2\gamma}\frac{|y_{1}-y_{2}|}{y_{2}^{ \gamma }}%
|\varphi^{\prime}(z)|M,
\end{align*}
where the constant $z\in (y_{1},y_{2})$ is produced by the Mean-Value Theorem. For arbitrary $t\in (y_{1},y_{2})$, we have
\begin{align*}
|\varphi^{\prime}(t)|  &  =|(1+\gamma )t^{ \gamma }(1-t)^{\gamma }- \gamma t^{1+\gamma }(1-t)^{\gamma-1}|\\
&  \leq(1+\gamma )y_{2}^{ \gamma }+  \gamma y_{2}^{1+ \gamma }2^{1-\gamma}.
\end{align*}
Hence,%
\[
III\leq2^{2\gamma}\big(1+ \gamma+ \gamma2^{\gamma-1}\big)M|y_{1}-y_{2}|.
\]
Combining the estimates for $I$, $II$, and $III$ gives a uniform (in $f$) Lipschitz constant $L$ such that %
\[
|T(f)(y_{2})-T(f)(y_{1})|\leq L|y_{2}-y_{1}|,\quad \forall 0<y_{1}<y_{2}\leq\delta,\quad \forall f\in \sX.
\]
Using (\ref{limit at 0}), it
follows that this inequality holds also for $y_{1}=0$. Because $L$ is independent of $f\in \sX$,  $T$
is equi-continuous.

\noindent\textbf{Step 5/6: }The family $\{T(f):\,f\in \sX\}$ is
equi-bounded (step 3) and equi-continuous (step 4). Hence, by the Ascoli--Arzel\'{a} theorem,
$T$ is compact. Then, the Schauder fixed point
theorem produces a function $f\in \sX$ such that $f(y)=T(f)(y)$ for all
$y\in\lbrack0,\delta]$. In particular, $f(0)=T(f)(0)=\gamma $, and
\begin{align}\label{int_rep_f}
f(y)=\frac{1}{\varphi(y)}\int_{0}^{y}\varphi(t)\left( a_0(t)+\frac{a(t)}{1-t}
f(t)^2\right)  dt,\quad y\in [0,\delta].
\end{align}
Since the right-hand side is continuously differentiable for $y>0$, it follows
that $f$ is continuously differentiable for $y\in(0,\delta)$. We need to show that $f$ is continuously  differentiable at $y=0$. Let $\phi$ be the function in Lemma \ref{lemma_phi}. Then,
\[
f(y)=(1+\gamma )  \gamma \phi(y)+\frac{1}{\varphi(y)}\int_{0}^{y}%
\varphi(t)\frac{a(t)}{1-t}f(t)^2\,dt,\quad y\in(0,\delta].
\]
Because $\phi\in \sC^1([0,\frac12])$, it remains to show that the second term on the right-hand side is also in $\sC^{1}([0,\delta])$. To this end, we define
\[
\omega(y):=\frac{1}{\varphi(y)}\int_{0}^{y}\varphi(t)\frac{a(t)}{1-t}f(t)^2\,dt,\quad y\in(0,\delta].
\]
Then, 
\begin{align*}
\omega^{\prime}(y)  &  =-\frac{\varphi^{\prime}(y)}{\varphi(y)^2}\int%
_{0}^{y}\varphi(t)\frac{a(t)}{1-t}%
f(t)^2dt+\tfrac{a(y)}{1-y}f(y)^2,\quad y\in(0,\delta), 
\end{align*}
which is continuous.  L'Hopital's rule and the definition of $\varphi(y)$ in \eqref{phi} ensure that the integral term in $\omega'(y)$ has a finite limit as $y\downarrow 0$. The continuity of $a(y)$ and $f(y)$ at $y=0$ ensures that the second term in $\omega'(y)$ also has a finite limit as $y\downarrow0$. Therefore, $\omega\in \sC^{1}([0,\delta])$.

\noindent\textbf{Step 6/6: } The local solution $f(y)$ for $y\in[0,\delta]$ is necessarily strictly positive. This is  because $f(0)=\gamma >0$ and should there be a point $y_0\in(0,\delta)$ with $f(y_0)=0$, then $f'(y_0) =  \tfrac{\gamma(1+\gamma)}{y_0}>0$, and so $f$ can never become zero. 
 
$\endproof$

\begin{remark}\label{rmk1} Because the solution in Theorem \ref{theorem local existence} belongs to $\sC^1([0,\delta])$, we know that $f(y) = \gamma +\int_0^y f'(t)dt$ for all $y\in[0,\delta]$. However, because the coefficient functions in \eqref{ode f}  have singularities, we cannot use \eqref{ode f} to write $\int_0^y f'(t)dt$ as a sum of three individual integrals. For example, the definition of $a_0(y)$ in \eqref{a0a1} implies that the first term satisfies $\int_0^y a_0(t) dt = \infty$ for all $y>0$.

\end{remark}

\begin{theorem}[Comparison]\label{Thm:Comparison} Let $\delta \in (0,1]$ and let $a,b:[0,\delta]\mapsto \R$ be continuous functions with $a(y)\le b(y)$ for all $y\in[0,\delta]$. Assume $f,g:[0,\delta]\mapsto \R$ solve the Cauchy problems
\begin{align*}
\begin{cases}
f^{\prime}(y)  =a_0(y)+\tfrac{a_1(y)}{1-y}f(y)+\tfrac{a(y)}{1-y}f(y)^2,\quad y\in (0,\delta),\\
f(0) =\gamma,
\end{cases}
\end{align*}
and 
\begin{align*}
\begin{cases}
g^{\prime}(y)  =a_0(y)+\tfrac{a_1(y)}{1-y}g(y)+\tfrac{b(y)}{1-y}g(y)^2,\quad y\in (0,\delta),\\
g(0) =\gamma.
\end{cases}
\end{align*}
Then, $f(y)\le g(y)$ for all $y\in[0,\delta]$. Furthermore, if $a(y)< b(y)$ for all $y\in[0,\delta]$, then $f(y)< g(y)$ for all $y\in(0,\delta]$.
\end{theorem}
\proof We argue by contradiction and assume there exists $y_1 \in (0,\delta]$ such that
$f(y_1) > g(y_1)$. Then, the function $F(y):= f(y)- g(y)$ satisfies $F(0) =0$ and $F(y_1) >0$. We define
$$
y_0 := \sup \{ y \in [0,y_1]: F(y) =0\} \in [0,y_1].
$$
Because $F$ is continuous, we have $y_0 \in [0,y_1)$, $F(y_0) = 0$, and $F(y) >0$ for all $y\in (y_0,y_1]$. For $y \in (y_0,y_1)$, we have 
\begin{align*}
F^{\prime}(y)  &=\frac{a_1(y)}{1-y}F(y)+\frac{a(y)}{1-y}f(y)^2-\frac{b(y)}{1-y}g(y)^2\\
&\le\frac{2\gamma +1}{1-y}F(y)+\frac{b(y)}{1-y}\big(f(y)^2-g(y)^2\big)\\
&=\frac{1}{1-y}F(y)\Big(2\gamma +1+b(y)\big(f(y)+g(y)\big)\Big)\\
&\le \frac{M}{1-y}F(y),\quad M:=\max_{y\in[y_0,\delta]} \Big(2\gamma +1+b(y)\big(f(y)+g(y)\big)\Big).
\end{align*}
Because $F(y_0)=0$, Gronwall's inequality yields the contradiction
$$
F(y) \le F(y_0) e^{\int_{y_0}^y \frac{M}{1-q}dq} =0,\quad y \in (y_0,y_1).
$$

When $a<b$, the integral representation \eqref{int_rep_f} and $f\le g$ give
\begin{align*}
f(y)&=\frac{1}{\varphi(y)}\int_{0}^{y}\varphi(t)\left( a_0(t)+\tfrac{a(t)}{1-t}
f(t)^2\right)  dt\\
&<\frac{1}{\varphi(y)}\int_{0}^{y}\varphi(t)\left( a_0(t)+\tfrac{b(t)}{1-t}
f(t)^2\right)  dt\\
&\le\frac{1}{\varphi(y)}\int_{0}^{y}\varphi(t)\left( a_0(t)+\tfrac{b(t)}{1-t}
g(t)^2\right)  dt\\
&=g(y),\quad y \in (0,\delta],
\end{align*}
where we have used $f >0$.

$\endproof$

Proposition \ref{lem_appendix} in Appendix \ref{appA} contains additional properties of the solution of the ODE defined in \eqref{fprime0} below.

\begin{theorem}\label{Thm:ODEbound} Let $\gamma\in(0,1)$ and $a_3 \in (-\infty, -\gamma]$. Then, there exists a function $f\in \sC([0,1]) \cap \sC^1([0,1))$ with $f(y) \in [0,1]$ for $y\in[0,1]$ that satisfies 
 \begin{align}\label{fprime0}
\begin{split}
f(y) = \frac{1}{y^{1+\gamma}(1-y)^{\gamma }}\int_0^y&\Big(\gamma(1-\gamma)\big(x(1-x)\big)^{ \gamma  }+ a_3\frac{x^{1+\gamma }}{(1-x)^{1-\gamma}}f(x)^2\Big)dx,\quad y\in(0,1).
\end{split}
\end{align}
Furthermore, the function $f$ satisfies the ODE
\begin{align}\label{fprime}
\begin{split}
f'(y) & = a_0(y)+ \tfrac{a_1(y)}{1-y}f(y)+\tfrac{a_3}{1-y}f(y)^2,\quad y\in(0,1),
\end{split}
\end{align}
and the boundary conditions 
\begin{align}\label{fboundconds}
 f(0) = \gamma, \quad 
 f(1) =-\frac{\gamma}{a_3}\le 1.
\end{align}
Finally, when $a_3\in (-\infty, -\gamma)$, we have $f(y) \in [0,1)$ for $y\in[0,1]$ and when $a_3 \in [-1, -\gamma]$, we have $f(y) \ge \gamma$ for $y\in[0,1]$.
\end{theorem}

\proof \noindent{\bf Step 1/4 (Global existence):} We can rewrite  \eqref{fprime} as
\begin{align}\label{fprime2g}
\begin{split}
f'(y) & = a_0(y) +\tfrac{f(y)}{1-y}\big( a_1(y) +a_3f(y)\big),\quad y\in(0,1).
\end{split}
\end{align}
Theorem \ref{theorem local existence} produces local existence and so let $f$ be a local solution of \eqref{fprime}, which is strictly positive. To see that $f(y)$ exists and is uniformly bounded for $y\in[0,1)$, we argue by contradiction.  If $f$ were to oscillate at some point $y_\infty \in (0,1]$ in the sense
\[
\liminf_{y\uparrow y_{\infty}}f(x)<\limsup_{y\uparrow y_{\infty}}f(x)=\infty,
\]
then we could find an increasing sequence $(y_{n})_{n\in\mathbb{N}}$ of local
maxima such that $y_{n}\uparrow y_{\infty}$, $f^{\prime}(y_{n})=0$, and
$f(y_{n})\rightarrow\infty$. Because $a_{3}<0$ and $a_{1}(y_{\infty})\in\mathbb{R}$, we see from \eqref{fprime2g} that $\lim_{n\rightarrow\infty
}f^{\prime}(y_{n})=-\infty$, which is a contradiction.  Alternatively, if we have continuous explosion in the sense
\[
\lim_{y\uparrow y_{\infty}}f(y)=\infty,
\]
 we get  from \eqref{fprime2g} that $\lim_{y\uparrow y_{\infty}}f^{\prime}(y)=-\infty$, which is again a  contradiction because it implies 
that $f$ is decreasing near $y_\infty$ while $f(y_\infty) = \infty$. All in all, $f(y)$ exists and is uniformly bounded for $y\in [0,1)$.

 It remains to show that $\lim_{y\uparrow 1}f(y)$ exists and do this by considering oscillations. 
To this end, first assume there exists a sequence $(y_n)_{n\in\N}\subset (0,1)$ monotonically increasing to $y=1$ such that $f'(y_n) = 0$ for all $n\in \N$.  Because $f$ is bounded, we can extract a subsequence if necessary to ensure that $\ell := \lim_{n\to\infty} f(y_n)$ exists in $[0,\infty)$. The ODE in \eqref{fprime2g} yields
\begin{align}\label{fprime2ga}
\begin{split}
0&= \lim_{n\to\infty} \Big(a_0(y_n) +\frac{f(y_n)}{1-y_n}\big( a_1(y_n) +a_3 f(y_n)\big)\Big)\\
&=(1+\gamma )\gamma +  \lim_{n\to\infty}\frac{f(y_n)}{1-y_n}\Big( \gamma +a_3f(y_n)\Big),
\end{split}
\end{align}
which forces $\ell \in \{0,-\frac{\gamma}{a_3}\}$. To see $\ell=-\frac{\gamma}{a_3}$, we argue by contradiction and assume $\ell=0$. In that case, \eqref{fprime2ga} yields
\begin{align}\label{fprime2gb}
\begin{split}
\lim_{n\to\infty}\frac{\ell - f(y_n)}{1-y_n}=-\lim_{n\to\infty}\frac{ f(y_n)}{1-y_n}&=1+\gamma >0,
\end{split}
\end{align}
which is impossible because $f(y)$ is nonnegative for all $y\in[0,1)$. So should $f(y)$ oscillate as $y\uparrow 1$, the sequence of corresponding max and min values $f(y_n)$, $n\in \N$, converges to $-\frac{\gamma}{a_3}$. Because $f(y)$ is uniformly bounded for $y\in[0,1)$, the alternative is that $f(y)$ is monotone as $y\uparrow 1$ in which case $\lim_{y\uparrow 1}f(y)$ obviously exists. All in all, this shows that $f\in\sC([0,1]) \cap \sC^1([0,1))$.

\noindent{\bf Step 2/4 (Boundary values):} By multiplying $(1-y)$ in \eqref{fprime}, we get
\begin{align}\label{fprime_take2}
(1-y)f'(y) & = (1-y)a_0(y)+ f(y)\big(a_1(y)+a_3f(y)\big),\quad y\in(0,1),
\end{align}
Because $f(1) = \lim_{y\uparrow 1}f(y)$, we see from \eqref{fprime_take2} that
\begin{align*}
 \lim_{y\uparrow 1}(1-y)f'(y) & =  f(1)\big(\gamma+a_3f(1)\big)=:c.
\end{align*}
To see \eqref{fboundconds}, we argue by contradiction. First, we assume $c>0$. For $\epsilon \in (0,c)$, we let $y_\epsilon \in (0,1)$ be such that
$$
\forall y \in (y_\epsilon,1): c-\epsilon \le (1-y)f'(y) \le c+\epsilon.
$$ 
For $y \in (y_\epsilon,1)$, we get 
\begin{align*}
f(y) - f(y_\epsilon) = \int_{y_\epsilon}^y f'(q)dq \ge (c-\epsilon) \int_{y_\epsilon}^y \frac1{1-q}dq.
\end{align*}
By using the Monotone Convergence Theorem and $f(1) <\infty$, this gives a contradiction. A similar argument rules out $c<0$.  Therefore, $c=0$ and so $f(1) \in \{0, -\frac{a_3}{\gamma}\}. $ To rule out $f(1) = 0$, we again argue by contradiction and assume $f(1) = 0$. The property 
$$
\lim_{y\uparrow 1} \big(a_1(y) + a_3 f(y)\big) = \gamma >0,
$$ 
allows us to find $y_0 \in (0,1)$ such that 
\begin{align}\label{posA}
\forall y\in(y_0,1): a_1(y) + a_3 f(y) >0.
\end{align}
Because $f >0$ and $a_0>0$, it follows from \eqref{fprime_take2} that $f'(y) >0$ for $y\in (y_0,1)$. This gives the contradiction
$$
0= f(1) = \lim_{y\uparrow 1}f(y)\ge f(y_0)>0.
$$
All in all, we have $f(1) = -\frac{a_3}{\gamma}$ as claimed in \eqref{fboundconds}.

\noindent\textbf{Step 3/4 (Global upper bound):} We claim that $f(y)\le 1$ for all
$y\in\lbrack0,1)$. Because $f$ is continuous on $[0,1]$, there exists $y_{1}\in\lbrack0,1]$ such
that%
\[
f(y_{1})=\max_{y\in\lbrack0,1]}f(y).
\]
If $y_{1}=0$, then $f(y)\leq f(0) = \gamma<1$ for all $y\in\lbrack0,1]$, and the claim
is proved. Similarly, if $y_{1}=1$, then $f(y)\leq f(1) = -\frac{\gamma}{a_{3}}\le1$ for
all $y\in\lbrack0,1]$. The only remaining case left to consider is $y_{1}%
\in(0,1)$ with $f^{\prime}(y_{1})=0$. In this case, the unique positive
solution of \eqref{fprime2g} with $f'(y_1)=0$ is
\begin{align}\label{eqsuperduper}
f(y_1)  &  =\frac{2a_{0}(y_1)(1-y_1)}{-a_{1}(y_1)+\sqrt{a_{1}(y_1)^2-4a_3 a_{0}(y_1)(1-y_1)}}.
\end{align}
and we claim that $f(y_1) \le 1$. This claim is proven by proving the inequality
\begin{align}\label{fineq1}
2a_{0}(y)(1-y)+a_{1}(y)\leq\sqrt{(a_{1}(y))^2-4a_{3}a_{0}(y)(1-y)},\quad y\in(0,1),
\end{align}
Assume there is $y\in(0,1)$ such that (otherwise there is nothing to prove)
$$
0\le 2a_{0}(y)(1-y)+a_{1}(y) = \frac{\gamma +(2 \gamma^2-1) (1-y)}{y}.
$$
Squaring both sides of  \eqref{fineq1}  produces the requirement
\begin{align}\label{fineq2}
0\ge y \left(a_3-\gamma ^2+\gamma +1\right)+\gamma ^2-1= y (a_3+\gamma ) - (1-\gamma ^2) (1-y).
\end{align}
The inequality \eqref{fineq2} holds because $a_3 \le -\gamma$.

When $a_3 \in (-\infty, -\gamma)$, we have $f(1)<1$. Then, for any maximizer $y_{1}\in(0,1)$ with $f^{\prime}(y_{1})=0$, it suffices to prove that the inequality in \eqref{fineq1} is strict. This follows because the left-hand-side of \eqref{fineq2} is strictly negative for all $y \in (0,1)$.

\noindent\textbf{Step 4/4 (Global lower bound):} Let $a_3 \in [-1, -\gamma]$. To show $f\ge \gamma$, we argue by contradiction and assume there is a point $y_1\in[0,1]$ such that $f(y_1)<\gamma$ and $f'(y_1) =0$. Because $f(0) =\gamma$ and $f(1) = - \frac{\gamma}{a_3} \ge \gamma$, it must be that  $y_1\in(0,1)$. The unique positive solution of $f'(y_1) =0$ is as in \eqref{eqsuperduper}. The contradictory property $f(y_1) \ge\gamma$ is equivalent to 
\begin{align}\label{Need2Q}
2a_{0}(y_1)(1-y_1)+\gamma a_{1}(y_1)\geq\gamma \sqrt{a_{1}(y_1)^2-4a_3a_{0}(y_1)(1-y_1)}.
\end{align}
The left-hand-side of \eqref{Need2Q} equals
$$
2a_{0}(y_1)(1-y_1)+\gamma a_{1}(y_1) = \frac{\gamma  (\gamma +1-y_1)}{y_1}>0.
$$
Squaring both sides of  \eqref{Need2Q}  produces the equivalent property
 $a_3+1\ge0$, which holds by assumption.  $\endproof$

\begin{remark}\label{rmkA} For given constants $y_0 \in (0,1)$ and $f_0 \in (0,1)$, we can argue as in the above
proof of Theorem \ref{Thm:ODEbound} and construct $f \in \sC([y_0,1]) \cap \sC^1([y_0,1))$ with $f(y) \in [0,1]$ for $y\in[y_0,1]$ that satisfies the ODE
 \begin{align}\label{fprime0take2}
\begin{split}
f'(y) & = a_0(y)+ \tfrac{a_1(y)}{1-y}f(y)+\tfrac{a_3}{1-y}f(y)^2,\quad y\in(y_0,1),
\end{split}
\end{align}
and the boundary conditions 
\begin{align}\label{fboundcondstake2}
 f(y_0) = f_0, \quad 
 f(1) =-\frac{\gamma}{a_3}.
\end{align}
Indeed, since the right-hand-side of \eqref{fprime0take2} is locally Lipschitz for $y \in [y_0,y_0+\delta_0]$ with $y_0 \in (0,1-\delta_0)$, standard existence theory for ODEs gives a local solution $f(y)$ for $y\in [y_0,y_0+\delta]$ for some $\delta \in (0,\delta_0)$. By reasoning as in Step 1, we see that $f(y)$ exists for $y\in [y_0,1)$ and is uniformly bounded. Step 2 still holds. Step 3 holds by using $f(y_0) = f_0 <1$ in place of $f(0) = \gamma <1$. 

Finally, assuming that $f_0 \in [\gamma,1)$ and $a_3 \in [-1,-\gamma]$, Step 4 still produces $f(y) \ge \gamma$ for all $y\in [y_0,1]$.
$\endproof$

\end{remark}

\subsection{Local ODE existence for all $\xi$}

For constants $\sigma_D^2, \xi, A$, and a continuous function $h:[0,1]\mapsto \R$, we define the coefficient functions (consistent with \eqref{a0a1})
\begin{align}\label{a0a1a2}
\begin{split}
a_0(y)&:= \frac{\gamma (1+\gamma )}{y},\quad y \in (0,1], \\
 a_1(y)& :=  \frac{(2 \gamma +1) y-(1+\gamma)}{y},\quad y \in (0,1],\\
a_2(h,y)& := \frac{\xi}{\sigma_D^2}e^{\int_0^y \frac{h(q)-1}{1-q}dq}-A,\quad y \in (0,1).
\end{split}
\end{align}

\begin{theorem}\label{ODE_local_existence} Let $\gamma \in (0,1)$, $\sigma_D^2>0$, and $A>1$.  For $\xi_0>0$, there exists $\delta_0\in (0,\frac12]$ such that 
\begin{enumerate}

\item For all $\xi>0$ with $|\xi - \xi_0| \le 1$, the integral equation 
  \begin{align}\label{z1a}
\begin{split}
h(y) &= \frac{1}{y^{1+\gamma }(1-y)^{\gamma}}\int_0^y\Big(a_0(x) (1-x)^{\gamma}+ \frac{a_2(h,x)}{(1-x)^{1-\gamma}}h(x)^2\Big)x^{1+\gamma}dx,\quad y\in(0,1),
\end{split}
\end{align}
 has a local unique solution $h_\xi\in \sC^1([0,\delta_0])$ and $h_\xi$ satisfies the  boundary value problem
 \begin{align}\label{wODE}
\begin{cases}
h'(y)  =a_0(y)+ \frac{a_1(y)}{1-y}h(y)+\frac{a_2(h,y)}{1-y}h(y)^2,\quad y\in(0,\delta_0), \\
h(0) = \gamma.
\end{cases}
\end{align}

\item For all $\xi_i>0$ with $|\xi_i - \xi_0| \le 1$ for $i \in \{1,2\}$,  comparison holds for  \eqref{wODE}  in the sense that $0<\xi_1 < \xi_2$ implies $h_{\xi_1}(y)<  h_{\xi_2}(y)$ for all $y\in(0,\delta_0]$, 


\end{enumerate}

\end{theorem}

\proof 1. The proof of local existence only requires minor modifications to the proof of Theorem \ref{theorem local existence}. For $0<\delta\leq\frac{1}{2}$ and $R>0$ with $R+\gamma \le 1$,  let $f\in\sX$ with $\sX$ defined in \eqref{sX}. We modify the operator $T$ in \eqref{Tf} by  defining
\begin{align}\label{Tf2}
T(f)(y):=\frac{1}{\varphi(y)}\int_{0}^{y}\varphi(t)\left(  a_0(t)+\frac{a_2(f,t)}{1-t}f(t)^2\right)  \,dt,\quad y\in (0,\delta].
\end{align}
To replace the estimates in \eqref{500} and \eqref{502}, we first need the bound
\begin{align}\label{500new}
\begin{split}
| a_2(f,t) |&= \Big|\tfrac{\xi}{\sigma_D^2} (1-t)e^{\int_0^t \frac{f(q)}{1-q}dq}-A\Big|\\
&\le (1+\tfrac{\xi_0}{\sigma_D^2}) (1-t)e^{(R+\gamma)\int_0^t \frac{1}{1-q}dq}+A\\
&\le (1+\tfrac{\xi_0}{\sigma_D^2}) (1-t)^{1-R-\gamma}+A\\
&\le 1+\tfrac{\xi_0}{\sigma_D^2} +A,\quad t \in [0,\delta].
\end{split}
\end{align}
 Therefore, 
\begin{align}\label{5000}
| a_2(f,t) |f(t)^2\le \big(1+\tfrac{\xi_0}{\sigma_D^2} +A\big) (R+\gamma)^2\le 1+\tfrac{\xi_0}{\sigma_D^2} +A,\quad t \in [0,\delta].
\end{align}
The estimate in \eqref{5000} implies that the analogous of  \eqref{500} and \eqref{502} hold with $M$ replaced by $1+\tfrac{\xi_0}{\sigma_D^2} +A$. With these modifications, the rest of the local existence proof is identical. 

We start by proving uniqueness for $y \in [0,y_1]$ for some $y_1 \in (0,\delta_0]$. To this end, we let $h$ and $\tilde{h}$ be two solutions of \eqref{z1a} and set $g:= h-\tilde h$. Because $h(0) = \tilde h (0) = \gamma$,  it suffices to prove $g(y)=0$ for $y\in[0,y_1]$. Moreover,  because $\gamma <1$, by taking $y_1 \in (0,1)$ small enough, we can assume $h(y)\le 1$ and $\tilde h(y) \le 1$ for $y \in [0,y_1]$.  For $y\in(0,y_1]$, we have
\begin{align}\label{uniq1}
\begin{split}
|g(y)| & \le \frac{1}{y^{1+\gamma }(1-y)^{\gamma}}\int_0^y\frac{x^{1+\gamma}}{(1-x)^{1-\gamma}}\Big| a_2(h,x)h(x)^2-a_2(\tilde h,x)\tilde h(x)^2\Big|dx\\
& \le \frac{1}{y^{1+\gamma }(1-y)^{\gamma}}\int_0^y\frac{x^{1+\gamma}}{(1-x)^{1-\gamma}}\Big| a_2(h,x)-a_2(\tilde h,x)\Big|h(x)^2dx\\
&+ \frac{1}{y^{1+\gamma }(1-y)^{\gamma}}\int_0^y\frac{x^{1+\gamma}}{(1-x)^{1-\gamma}}|a_2(\tilde h,x)|\Big| h(x)^2-\tilde h(x)^2\Big|dx.
\end{split}
\end{align}
We consider the last two integrals in \eqref{uniq1} separately. The Mean-Value Theorem gives the inequality
$$
|e^a - e^b| \le e^{|a|+|b|}|a-b|,\quad a,b\in \R.
$$
Therefore, the first integral in \eqref{uniq1} is bounded by
\begin{align}\label{uniq11}
\begin{split}
& \frac{1}{y^{1+\gamma }(1-y)^{\gamma}}\int_0^y\frac{x^{1+\gamma}}{(1-x)^{1-\gamma}}\frac{\xi(1-x)}{\sigma_D^2}\Big| e^{\int_0^x \frac{h(q)}{1-q}dq}-e^{\int_0^x \frac{\tilde h(q)}{1-q}dq}\Big|dx\\
&\le \frac{\xi}{\sigma_D^2(1-y_1)^{\gamma}}\int_0^ye^{\int_0^x \frac{h(q)+\tilde h(q)}{1-q}dq}\int_0^x \frac{|g(q)|}{1-q}dq dx\\
&\le \frac{\xi }{\sigma_D^2(1-y_1)^{\gamma}}\int_0^y\frac1{(1-x)^2}\int_0^y \frac{|g(q)|}{1-q}dq dx\\
&\le \frac{\xi}{\sigma_D^2(1-y_1)^{\gamma+2+1}}\int_0^y |g(q)|dq.
\end{split}
\end{align}

For the second integral in \eqref{uniq1}, the estimate in \eqref{500new} ensures that $a_2(\tilde h,x)$ is uniformly bounded by some constant $c_1>0$.  Therefore, 
\begin{align}\label{uniq12}
\begin{split}
&\frac{1}{y^{1+\gamma }(1-y)^{\gamma}}\int_0^y\frac{x^{1+\gamma}}{(1-x)^{1-\gamma}}|a_2(\tilde h,x)|\big| h(x)^2-\tilde h(x)^2\big|dx\\
&\le \frac{c_1}{(1-y_1)}\int_0^y\big(h(x)+\tilde h(x)\big)\big|h(x)-\tilde h(x)\big|dx\\
&\le \frac{2 c_1}{(1-y_1)}\int_0^y\big|g(x)\big|dx.
\end{split}
\end{align}

By combining \eqref{uniq1} with the two estimates \eqref{uniq11} and \eqref{uniq12}  we get
$$
|g(y) | \le c_2 \int_0^y\big|g(x)\big|dx,\quad y\in[0,y_1],
$$
for some constant $c_2>0$. Because $g(0) = 0$, Gronwall's inequality produces $g(y)=0$ for all  $y\in[0,y_1]$.

Next, we prove uniqueness for $y\in [y_1,\delta_0]$. The function  $ [y_1,\delta_0] \ni y \mapsto \int_0^y \frac{h(q)}{1-q}dq$ transforms the ODE \eqref{wODE} into a second-order ODE. Because $y_1>0$, there is no singularity and standard uniqueness arguments apply.

\noindent 2. To argue by contradiction, we assume there exists $y_0\in (0,\delta_0]$ such that $h_1(y_0)\ge  h_2(y_0)$. First, because $ \xi_1 < \xi_2$, there exists $y_1 \in (0,1)$ such that
$$
\forall y \in [0,y_1]: \xi_1(1-y)e^{\int_0^y \frac{h_1(q)}{1-q}dq} < \xi_2(1-y)e^{\int_0^y \frac{ h_2(q)}{1-q}dq}.
$$
The strict comparison in Theorem \ref{Thm:Comparison} gives $h_1(y)< h_2(y)$ for $y\in [0,y_1]$. 
Second, we define
$$
y_2 := \sup\{ y\in [0,\delta_0]:  \forall q\in (0,y]\;h_1(q) < h_2(q)\} \in [y_1,y_0].
$$
The continuity of $h_1$ and $h_2$ gives $h_1(y_2) = h_2(y_2)$. Because $\xi_1 < \xi_2$, we have
$$
\forall y \in [0,y_2]: \xi_1(1-y)e^{\int_0^y \frac{h_1(q)}{1-q}dq} < \xi_2(1-y)e^{\int_0^y \frac{h_2(q)}{1-q}dq}.
$$
The strict comparison in Theorem \ref{Thm:Comparison} gives the contradiction  $h_1(y)< h_2(y)$ for $y\in (0,y_2]$. 
$\endproof$

\subsection{Global ODE existence for $\xi$ small}

\begin{theorem}\label{ODE_existence} Let $\gamma \in (0,1)$, $\sigma_D^2>0$, and $A>1$. For a constant $\xi$ satisfying
\begin{align}\label{xiass}
0\le \frac{\xi}{\sigma_D^2}  <A-\gamma,
\end{align}
 the integral equation   \eqref{z1a} has a  global solution $h_\xi\in \sC([0,1])\cap \sC^1([0,1))$ with $0< h_\xi \le 1$. Furthermore,  $h_\xi$ satisfies the boundary value problem \eqref{wODE} globally, as well as the boundary condition
 \begin{align}\label{c0}
 h(1)=c_0,\;\; c_0:= \frac{\gamma}A \in (0,\gamma).
 \end{align}
\end{theorem}

\proof  Let $f$ be as in Theorem \ref{Thm:ODEbound} for the constant
\begin{align}\label{a2}
a_{3}  &:=\frac{\xi}{\sigma_D^2}-A< -\gamma <0.
\end{align}

\noindent {\bf Step 1/3  (Global Existence):} Let $[0,\delta_1)$ be the maximal interval of existence for $h = h_\xi$ produced by Theorem \ref{ODE_local_existence}(1). First, we claim $h(y) \le 1$ for $y \in [0,\delta_1)$. To see this, we argue by contradiction and assume there exists $y_1\in [0,\delta_1)$ such that $h(y_1)>1$. Because $h(0) = \gamma<1$, it must be that $y_1\in (0,\delta_1)$. We define
$$
y_0 := \sup\{ y\in [0,y_1): h(y) \le 1\} \;\in (0,y_1).
$$
The continuity of $h$ gives
$$
 \forall y\in [0,y_0]:  h(y) \le 1\quad  \text{and}\quad  h(y_0) =1.
$$
Therefore, for $y\in [0,y_0]$, we have
\begin{align}\label{boundddd1}
a_2(h_{\xi},y) &= \frac{\xi}{\sigma_D^2}e^{\int_0^y \frac{h(q)-1}{1-q}dq} - A\le   \frac{\xi}{\sigma_D^2} - A=a_3.
\end{align}
Theorem \ref{Thm:ODEbound} gives an ODE solution $f$ with $f< 1$. The comparison result in Theorem \ref{Thm:Comparison} gives the contradiction 
$$
1=h(y_0) \le f(y_0) <1.
$$

Because $h(y) \le 1$ for all $y\in[0,\delta_1)$, \eqref{boundddd1} holds and we can again use 
Theorems  \ref{Thm:Comparison}  and \ref{Thm:ODEbound} to see
$$
h(y) \le f(y) < 1,\quad y\in[0,\delta_1).
$$
Therefore, $\delta _1 = 1$.

\noindent {\bf Step 2/3 (Boundary values):} We start by writing
\[
(1-y)e^{\int_{0}^{y}\frac{h(x)}{1-x}\,dx}=e^{\int_{0}^{y}\frac{h(x)-1}%
{1-x}\,dx}.
\]
Since $h\leq1$, the integral $\int_{0}^{1}\frac{h(x)-1}{1-x}\,dx$ exists in
$[-\infty,0]$, and so the following limit exists 
\begin{align}\label{Lstar}
L:=\lim_{y\uparrow1}(1-y)e^{\int_{0}^{y}\frac{h(x)}{1-x}\,dx}=e^{\int_{0}%
^{1}\frac{h(x)-1}{1-x}\,dx}\in\lbrack0,1].
\end{align}

As in Step 1 in the proof of Theorem \ref{Thm:ODEbound}, assume first that there exists a sequence $(y_{n})_{n\in\mathbb{N}}\subset(0,1)$ monotonically increasing to $y=1$ such that $h^{\prime}(y_{n})=0$ for all
$n\in\mathbb{N}$. Because $h\le 1$, we can extract a subsequence, if
necessary, to ensure that $\ell:=\lim_{n\rightarrow\infty}h(y_{n})\in
\lbrack0,1]$ exists. The ODE in \eqref{wODE} yields
\begin{align*}
0  & =\lim_{n\rightarrow\infty}\left(  a_{0}(y_{n})+\frac{h(y_{n})}{1-y_{n}%
}\big(a_{1}(y_n)+a_{2}(h,y)h(y_{n})\big)\right)  \\
& =(1+\gamma)\gamma+\lim_{n\rightarrow\infty}\frac{h(y_{n})\big(a_{1}(y_{n}%
)+a_{2}(h,y_{n})h(y_{n})\big)}{1-y_{n}}.
\end{align*}
This implies that%
\begin{align*}
0&=\lim_{n\rightarrow\infty}h(y_{n})\big(a_{1}(y_{n})+a_{2}(h,y_{n})h(y_{n})\big)\\
&=\ell\left(  \gamma+ \Big( \frac{\xi}{\sigma_D^2}L -A\Big)\ell\right)  .
\end{align*}
Hence, either $\ell=0$ or $ \ell=\ell_{1}$ where
\begin{equation}
\ell_1:=\frac{\gamma}{A-\frac{\xi}{\sigma_D^2} L}.\label{ell}%
\end{equation}
We argue by contradiction to rule out $\ell=0$. We assume $\ell=0$ to get
\begin{align*}
0  & =\lim_{n\rightarrow\infty}\left(  a_{0}(y_{n})+\frac{h(y_{n})}{1-y_{n}%
}\big(a_{1}(y_{n})+a_{2}(h,y_{n})h(y_{n})\big)\right)  \\
& =(1+\gamma)\gamma+\lim_{n\rightarrow\infty}\frac{h(y_{n})}{1-y_{n}}\gamma.
\end{align*}
This is a contradiction because $h$ is positive.

Based on the above, should $h$ oscillate as $y\uparrow1$, the sequence of corresponding max and min values $h(y_{n})$ would converge to $\ell_{1}$ given in (\ref{ell}). Therefore, $\ell:= \lim_{y\uparrow1}h(y)$ exists and equals  $\ell_{1}$. Because $h\le1$, the alternative is that $h$ is monotone near $y=1$, in which case the limit $\ell:=\lim_{y\uparrow1}h(y)$ exists. To show $\ell=\ell_{1}$ when $h$ is monotone near $y=1$, we follow the argument in Step 2 in the proof of Theorem \ref{Thm:ODEbound} but replace \eqref{fprime_take2} with
\begin{align*}
(1-y)h'(y) & = (1-y)a_0(y)+ h(y) a_1(y)+a_2(h,y)h(y)^2,\quad y\in(0,1).
\end{align*}
Taking limits yields
\begin{align*}
\lim_{y\uparrow 1}(1-y)h'(y) & = \ell \gamma+\Big( \frac{\xi}{\sigma_D^2} L -A\Big)\ell^2.
\end{align*}
As in the proof of  Theorem \ref{Thm:ODEbound}, we get $ \ell \gamma+\Big( \frac{\xi}{\sigma_D^2} L -A\Big)\ell^2 =0$ and so either $\ell=0$ or $\ell=\ell_{1}$ for $\ell_{1}$ in (\ref{ell}). We rule out $\ell=0$ as in Step 2 of the proof of Theorem \ref{Thm:ODEbound}. Therefore, when $h(y)$ is monotone near $y=1$, we also have $\ell =\ell_1$.

Let us show that $\ell=c_0$ where $c_0$ is from \eqref{c0}. Since $\frac{\xi}{\sigma_D^2} < A -\gamma$ and $L\in[0,1]$, we have that $\ell_1$ from \eqref{ell} satisfies
\begin{align}\label{l1bound1}
\ell_{1}=\frac{\gamma}{A-\frac{\xi}{\sigma_D^2} L}\leq1.
\end{align}
We consider two cases. First, if $\ell_{1}<1$, then taking $\varepsilon>0$ so small
that $\eta:=\ell_{1}+\varepsilon<1$, there exists $y_{1}\in(0,1)$ such that%
\[
\forall y\in\lbrack y_{1},1]: h(y)\leq\eta<1.
\]
It follows that $h(y)\leq\eta$ for all
$y\in\lbrack y_{1},1]$, while $h(y)\leq1$ for all $y\in\lbrack0,y_{1}]$. Hence, for all $y\in\lbrack y_{1},1]$,%
\begin{align*}
0 &  \leq(1-y)e^{\int_{0}^{y}\frac{h(q)}{1-q}dq}\\
&\leq(1-y)e^{\int_{0}^{y_{1}%
}\frac{1}{1-q}dq}e^{\int_{y_{1}}^{y}\frac{\eta}{1-q}dq}\\
&=\frac{1-y}{1-y_{1}%
}\left(  \frac{1-y_{1}}{1-y}\right)  ^{\eta},
\end{align*}
which converges to zero as $y\uparrow1$. Therefore, when $\ell_1<1$, we have $L=0$
and so $\ell=c_0$.

Second, assume $\ell_{1}=1$. From \eqref{ell}, we see that $\ell_{1}=1$ implies 
\[
\frac{\xi L}{\sigma_D^2}=A-\gamma.
\]
Therefore, the upper bound in \eqref{xiass} gives $L>1$, which contradicts \eqref{Lstar}. Thus, $\ell_{1}=1$ cannot happen and $\ell=c_0$.

To see that $h'\in \sC([0,1))$, we can repeat the argument from Step 2 of the proof of Theorem \ref{theorem local existence}.

\noindent {\bf Step 3/3:} To see $h(y)>0$ for all $y\in[0,1]$, the boundary values \eqref{wODE} imply that it suffices to show that there is no point $y\in (0,1)$ such that 
\begin{align}\label{hhprime}
h(y) = h'(y) = 0.
\end{align}
However, the ODE \eqref{wODE} makes \eqref{hhprime} impossible because $a_0(y) >0$ for all $y\in[0,1]$.

$\endproof$

\subsection{Explosion for $\xi$ big}
\begin{lemma}\label{Ricexplosion}  Let $\gamma \in (0,1)$, $\sigma_D^2>0$, and $A>1$.  For $y_0 \in (0,1)$, there exists $\xi(y_0)\in (0,\infty)$ such that 
$$
\forall \xi \ge \xi(y_0)\;:\; h_\xi(y_0) =\infty.
$$ 
\end{lemma}
\proof Let $y_0 \in (0,1)$ be arbitrary and assume to the contrary that for all $\xi>0$, the local solution $h_\xi(y)$ of \eqref{wODE} exists for $y\in[0,y_0]$ and that $\lim_{\xi\uparrow \infty} h_\xi(y_0) <\infty$ (this limit exists because Theorem \ref{ODE_local_existence}  ensures that $h_\xi(y_0)$ is increasing in $\xi$). We consider $\xi>0$ such that 
\begin{align}\label{compa0}
\frac{\xi}{\sigma_D^2}> \frac{A}{1-y_0}.
\end{align}

By \eqref{z1a}, the ODE for $g_\xi(y) := h_\xi(y) (1-y)^\gamma y^{1+\gamma}$ gives 
\begin{align}\label{compa1}
\begin{split}
g'_\xi(y) &= a_0(y)(1-y)^\gamma y^{1+\gamma}+  \tfrac{a_2(h_\xi,y)}{\big((1-y)y\big)^{1+\gamma}}g_\xi(y)^2\\
&= \gamma(1-\gamma)(1-y)^\gamma y^\gamma+  \tfrac{\frac{\xi}{\sigma_D^2} e^{\int_0^y \frac{h_\xi(q)-1}{1-q}dq}- A}{\big((1-y)y\big)^{1+\gamma}}g_\xi(y)^2\\
&\ge \gamma(1-\gamma)(1-y)^\gamma y^\gamma+  \tfrac{\frac{\xi}{\sigma_D^2} (1-y)-A}{\big((1-y)y\big)^{1+\gamma}}g_\xi(y)^2\\
&\ge \gamma(1-\gamma)(1-y)^\gamma y^\gamma+  \tfrac{\frac{\xi}{\sigma_D^2} (1-y_0)-A}{\big((1-y)y\big)^{1+\gamma}}g_\xi(y)^2
,\quad y\in(0,y_0),
\end{split}
\end{align}
where the first inequality uses $h_\xi>0$.  Integrating \eqref{compa1} and using \eqref{compa0} and $g_\xi(0)=0$ give
\begin{align}\label{compa15}
g_\xi(\tfrac{y_0}2) \ge  \gamma(1-\gamma) \int_0^{\frac{y_0}2} (1-q)^\gamma q^\gamma dq>0.
\end{align}

To create a contradiction, we define the constant
$$
c_\xi:= \tfrac{\frac{\xi}{\sigma_D^2} (1-y_0)- A}{\big((1-y_0)\frac{y_0}2\big)^{1+\gamma}}>0.
$$
Because $c_\xi$ is affine in $\xi$, we can increase $\xi>0$ if needed such that $\xi$ satisfies both \eqref{compa0} and
\begin{align}\label{compa2}
2 \le  c_\xi y_0\gamma(1-\gamma)\int_0^{\frac{y_0}2} (1-q)^\gamma q^\gamma dq.
\end{align}
The  Riccati equation 
\begin{align*}
\begin{cases}
f'(y) = c_\xi f(y)^2,\quad y> \tfrac{y_0}2,\\
f(\tfrac{y_0}2) = g_\xi(\tfrac{y_0}2),
\end{cases}
\end{align*}
has the explicit solution
$$
f(y) = \frac{ g_\xi(\tfrac{y_0}2) }{1-c_\xi g_\xi(\tfrac{y_0}2) (y-\tfrac{y_0}2)},\quad y \in[\tfrac{y_0}2, \tfrac1{c_\xi g_\xi(\tfrac{y_0}2)}+\tfrac{y_0}2).
$$
The solution $f$ explodes at 
\begin{align}\label{f_explosion_time}
y=\tfrac1{c_\xi g_\xi(\tfrac{y_0}2)}+\tfrac{y_0}2\le y_0,
\end{align}
where the inequality follows from \eqref{compa15} and \eqref{compa2}. For $y\in [\tfrac{y_0}2,y_0]$, \eqref{compa1} gives
\begin{align}\label{compa3}
\begin{split}
g'_\xi(y) &\ge  \tfrac{\frac{\xi}{\sigma_D^2} (1-y_0)- A}{\big((1-y)y\big)^{1+\gamma}}g_\xi(y)^2\ge  c_\xi g_\xi(y)^2.
\end{split}
\end{align}
Then, comparison gives us that $g_\xi(y)\ge f(y)$ for $y\ge\frac{y_0}2$. Therefore, $g_\xi$ also explodes at or before $y$ in \eqref{f_explosion_time}. 
$\endproof$

\subsection{Lipschitz estimates }\label{sec_growthrate}

For $\xi>0$, we denote by $h_\xi$ a local solution of the boundary value problem \eqref{wODE} and define the constant
\begin{align}\label{yxi}
y_\xi := \inf\{ y>0 : h_{\xi}(y) =1\} \land 1 \in (0,1],
\end{align}
where $\inf \emptyset := \infty$ and $a\land b := \min\{a,b\}$ for $a,b\in\R$. We also define the decreasing function
\begin{align}\label{keyy2A}
F_\xi(y) := \frac{\xi}{\sigma_D^2}e^{\int_0^y \frac{h_\xi(q)-1}{1-q}},\quad y \in [0,y_\xi].
\end{align}

\begin{lemma}\label{Lem_Lip}  Let $\gamma \in (0,1)$, $\sigma_D^2>0$, $A>1$, $y_0 \in (0,1)$, and $\bar\xi>0$. Then, there exist constants $M_1>0$ and $M_2>0$ such that
\begin{align}
|h_{\xi_1}(y)-h_{\xi_2}(y)|\leq M_1 y|\xi_1-\xi_2|,\quad y \in [0, y_{\xi_1}\land  y_{\xi_2} \land y_0],\quad \xi_1,\xi_2\in [0,\bar\xi], \label{Lipp1}\\
|F_{\xi_1}(y)-F_{\xi_2}(y)|\leq M_2 |\xi_1-\xi_2|,\quad y \in [0, y_{\xi_1}\land  y_{\xi_2} \land y_0],\quad \xi_1,\xi_2\in [0,\bar\xi].\label{Lipp2}
\end{align}

\end{lemma}

\proof  For  $\xi_1,\xi_2\in [0,\bar\xi]$,  we let $h_{1}$ and
$h_{2}$ be the corresponding local ODE solutions of  \eqref{wODE} 
\[
h_{i}^{\prime}(y)+\frac{1+\gamma}{y}\big(h_{i}(y)-\gamma\big)=\frac{\gamma}{1-y}h_{i}(y)+\frac{a_{2}(\xi_{i},h_{i},y)}{1-y}h_{i}(y)^2,\quad y\le y_{\xi_i},
\]
where we have augmented the $a_2$ function from \eqref{a0a1a2} by writing
\begin{align*}
a_{2}(\xi_{i},h_{i},y)  &:=\frac{\xi_{i}}{\sigma_D^2}e^{\int_{0}%
^{y}\frac{h_{i}(q)-1}{1-q}\,dq}-A,\quad y\le y_{\xi_i}.
\end{align*}
Subtracting and multiplying by $y^{1+\gamma}$ yield
\begin{align*}
&  y^{1+\gamma}\big(h_{1}^{\prime}(y)-h_{2}^{\prime}(y)\big)+y^{\gamma}(1+\gamma
)\big(h_{1}(y)-h_{2}(y)\big)\\
&=y^{1+\gamma}\frac{\gamma}{1-y}\big(h_{1}(y)-h_{2}(y)\big)\\
&  +y^{1+\gamma}\frac{a_{2}(\xi_{1},h_{1},y)-a_{2}(\xi_{2},h_{2},y)}{1-y}
h_{1}(y)^2+y^{1+\gamma}\frac{a_{2}(\xi_{2},h_{2},y)}{1-y}\big(h_{1}(y)^2-h_{2}(y)^2\big),
\end{align*}
for  $y \le  y_{\xi_1}\land  y_{\xi_2}$. Computing the $y$ derivative gives
\begin{align*}
&  \Big(  y^{1+\gamma}\big(h_{1}(y)-h_{2}(y)\big)\Big)'  =y^{1+\gamma
}\frac{\gamma}{1-y}\big(h_{1}(y)-h_{2}(y)\big)\\
&  +y^{1+\gamma}\frac{a_{2}(\xi_{1},h_{1},y)-a_{2}(\xi_{2},h_{2},y)}{1-y}h_{1}(y)^2+y^{1+\gamma}\frac{a_{2}(\xi_{2},h_{2},y)}{1-y}\big(h_{1}(y)^2-h_{2}(y)^2\big).
\end{align*}
Because $h_{i}\in \sC^1([0,y_{\xi_i}])$ and $h_i(0) = \gamma$, we can integrate from $0$ to $y$ with $y\in(0,  y_{\xi_1}\land  y_{\xi_2}]$ to get%
\begin{align}\label{lipsss1}
\begin{split}
&  y^{1+\gamma}\big(h_{1}(y)-h_{2}(y)\big)\\
&=\gamma\int_{0}^y\frac{t^{1+\gamma}}{1-t}\big(h_{1}(t)-h_{2}(t)\big)dt\\
&+\int_{0}^{y}t^{1+\gamma}\frac{a_{2}(\xi_{1}%
,h_{1},t)-a_{2}(\xi_{2},h_{2},t)}{1-t}h_{1}(t)^2\,dt  \\
&+\int_{0}^{y}a_{2}(\xi_{2},h_{2},t)\frac{t^{1+\gamma}}{1-t}\big(h_{1}(t)^2-h_{2}(t)^2\big)dt\\
&=\int_{0}^y\frac{t^{1+\gamma}}{1-t}\Big(\gamma+ a_{2}(\xi_{2},h_{2},t) \big(h_{1}(t)+h_{2}(t)\big)\Big)\big(h_{1}(t)-h_{2}(t)\big)dt\\
&+\int_{0}^{y}t^{1+\gamma}\frac{a_{2}(\xi_{1}%
,h_{1},t)-a_{2}(\xi_{2},h_{2},t)}{1-t}h_{1}(t)^2\,dt.
\end{split}
\end{align}

The bound $h_i(t) \le 1$ for  $t\in [0,y_{\xi_1}\land  y_{\xi_2}]$ gives
\begin{align*}
\left\vert \frac{a_{2}(\xi_{2},h_{2},t)}{1-t}\big(h_{1}(t)+h_{2}(t)\big)\right\vert\le 2\frac{\frac{\xi_2}{\sigma_D^2}+A}{1-y_{0}}\le 2\frac{\frac{\bar \xi}{\sigma_D^2}+A}{1-y_{0}}=:M_0,\quad t\in [0,y_{\xi_1}\land  y_{\xi_2}  \land y_0].
\end{align*}
Because
\[
a_{2}(\xi_{1},h_{1},t)-a_{2}(\xi_{2},h_{2},t)=\frac{(1-t)(\xi
_{1}-\xi_{2})}{\sigma_D^2}e^{\int_{0}^{t}\frac{h_{1}(q)}{1-q}\,dq}+\frac{(1-t)\xi_{2}}%
{\sigma_D^2}\left(  e^{\int_{0}^{t}\frac{h_{1}(q)}{1-q}\,dq}-e^{\int%
_{0}^{t}\frac{h_{2}(q)}{1-q}\,dq}\right),
\]
the Mean-Value Theorem applied to the function
$$
[0,1]\ni \theta\mapsto  e^{\int_{0}^{t}\frac{\theta
h_{1}(q)+(1-\theta)h_{2}(q)}{1-q}\,dq},
$$
gives $\theta(t)\in [0,1]$ such that
\begin{align}\label{MVT1}
\begin{split}
&  \frac{|a_{2}(\xi_{1},h_{1},t)-a_{2}(\xi_{2},h_{2},t)|}{1-t}h_{1}(t)^2\\
&\leq\frac{|\xi_{1}-\xi_{2}|}{\sigma_D^2}+\frac{\xi_{2}}{\sigma_D^2}\left\vert e^{\int_{0}^{t}\frac{\theta(t)
h_{1}(q)+(1-\theta(t))h_{2}(q)}{1-q}\,dq}\int_{0}^{t}\frac{h_{1}(q)-h_{2}%
(q)}{1-q}\,dq\right\vert \\
&  \leq\frac{|\xi_{1}-\xi_{2}|}{\sigma_D^2}+\frac{\xi_{2}}%
{\sigma_D^2(1-t)^2}\int_{0}^{t}|h_{1}(q)-h_{2}(q)|\,dq, \quad t\in [0,y_{\xi_1}\land  y_{\xi_2}  \land y_0].
\end{split}
\end{align}
Therefore, for $y\in[0,y_{\xi_1}\land  y_{\xi_2}  \land y_0]$, we have by \eqref{lipsss1}
\begin{align*}
y^{1+\gamma}|h_{1}(y)-h_{2}(y)| &  \leq\left(\frac{\gamma}{1-y_0} +M_0\right) \int%
_{0}^{y}t^{1+\gamma}|h_{1}(t)-h_{2}(t)|\,dt\\
&  +\frac{1}{\sigma_D^2}|\xi_{1}-\xi_{2}|\int_{0}^{y}t^{1+\gamma
}\,dt\\
&  +\frac{\xi_{2}}{\sigma_D^2(1-y_{0})^2}\int_{0}^{y}t^{1+\gamma}dt\int%
_{0}^{y}|h_{1}(q)-h_{2}(q)|\,dq.
\end{align*}
In turn, for $y\in[0,y_{\xi_1}\land  y_{\xi_2}  \land y_0]$, we have
\begin{align*}
|h_{1}(y)-h_{2}(y)| &  \leq\left(  \frac{\gamma}{1-y_0}+M_0+\frac{\xi_2%
y}{\sigma_D^2(1-y_{0})^2}\right)  \int_{0}^{y}|h_{1}(q)-h_{2}(q)|\,dq\\
&  +\frac{1}{\sigma_D^2}|\xi_{1}-\xi_{2}|y.
\end{align*}
For $y\in [0,y_{\xi_1}\land  y_{\xi_2}  \land y_0]$, Gronwall's inequality gives \eqref{Lipp1} because
\begin{align}\label{ineq 1}
\begin{split}
|h_{1}(y)-h_{2}(y)|&\leq\frac{|\xi_{1}-\xi_{2}|}{\sigma_D^2}ye^{\left(   \frac{\gamma}{1-y_0}+M_0+\frac{\xi_2}{\sigma_D^2(1-y_{0})^2}\right)  y}\\
&\leq M_1y|\xi_{1}-\xi_{2}|,\quad M_1:= \frac{1}{\sigma_D^2}e^{\left(\frac{\gamma}{1-y_0}+M_0+\frac{\bar\xi}{\sigma_D^2(1-y_{0})^2}\right)y_0}.
\end{split}
\end{align}

Finally, to see \eqref{Lipp2}, we use \eqref{MVT1} to see  for $y\in[0,y_{\xi_1}\land  y_{\xi_2}  \land y_0]$
\begin{align*}
|F_{\xi_1}(y)-F_{\xi_2}(y)|&\leq\frac{|\xi_{1}-\xi_{2}|}{\sigma_D^2}(1-y)+\frac{\xi_{2}}%
{\sigma_D^2}\int_{0}^{y}|h_{1}(q)-h_{2}(q)|\,dq\\
&\leq\frac{|\xi_{1}-\xi_{2}|}{\sigma_D^2}+\frac{\bar\xi}{\sigma_D^2} M_1|\xi_{1}-\xi_{2}|\int_{0}^{y}qdq\\
& = M_2|\xi_{1}-\xi_{2}|,\quad M_2:= \frac{1+\frac12\bar\xi M_1}{\sigma_D^2}.
\end{align*}

$\endproof$

The uniqueness claim in Theorem \ref{theorem_main_h} follows from the following result.

\begin{corollary} \label{cor_uniqueness} Let $\gamma \in (0,1)$, $\sigma_D^2>0$, and $A>1$. For all $\xi>0$, the boundary value problem  \eqref{wODE} has a unique local solution.
\end{corollary}

\proof Existence follows from Theorem \ref{ODE_local_existence}. To prove uniqueness, we let $h_i: I_i \to (0,\infty)$, $i \in \{1,2\}$,  be two solutions where $I_i \subseteq [0,1]$ is their maximal interval of existence. We let $y_i$ be the right-end point of $I_i$. Because $h_i(0) = \gamma<1$, there exists $y_0 \in (0,y_1 \land y_2)$ such that $h_i(y) <1$ for all $y \in [0,y_2]$.
By applying \eqref{Lipp1} with $\xi_1 := \xi_2:= \bar\xi:=\xi$, we see that $h_1(y) = h_2(y)$ for all $y\in [0,y_0]$. Because $y_0>0$, there is no singularity and standard uniqueness arguments imply $I_1 = I_2$ and $h_1(y) = h_2(y)$ for all $y \in I_1$.
$\endproof$

\subsection{Global ODE existence for critical $\xi$}
The existence claim in Theorem \ref{theorem_main_h} follows from the following result after we note that the ODE \eqref{wODE} coincides with the ODE in \eqref{KEY_ODE}. This property follows because the integral in \eqref{L0} below ensures
$$
(A-\gamma)e^{\int_y^1 \frac{1-h_{\xi_0}(q)}{1-q}dq} = \frac{\xi_0}{\sigma_D^2}  e^{\int_0^y \frac{h_{\xi_0}(q)-1}{1-q}dq},\quad y\in[0,1].
$$

\begin{lemma}\label{lemma:xi0}  Let $\gamma \in (0,1)$, $\sigma_D^2>0$, $A>1$, and define the set 
$$
\Xi:= \big\{ \xi >0 :  \text{$h_\xi\in \sC([0,1])\cap \sC^1([0,1))$ solves \eqref{wODE} with }  h_\xi(0) = \gamma \text{ and }h_\xi(1) \le  \gamma\big\}.
$$
Then, the following properties hold:

\begin{enumerate}

\item $\Xi$ is a non-empty and bounded subset of $(0,\infty)$.
\item For $\xi\in \Xi$, $h_\xi(y)<1$ for all $y\in[0,1]$.
\item  $\xi_0 := \sup \Xi \not\in \Xi$.  
\item $h_{\xi_0}\in  \sC([0,1])\cap \sC^1([0,1))$ solves \eqref{wODE} with $h_{\xi_0}(1)=1$.
\item $\int_0^1 \frac{1-h_{\xi_0}(q)}{1-q}dq\in (0,\infty)$ and the integral is given by
\begin{align}\label{L0}
&\lim_{y\uparrow 1} e^{\int_0^y \frac{h_{\xi_0}(q)-1}{1-q}dq} = \frac{\sigma_D^2(A-\gamma)}{\xi_0} \in (0,1).
\end{align}
\item $h_{\xi_0}(y)\ge \gamma$  for all $y\in[0,1]$.
\item $h_{\xi_0} \in  \sC^1([0,1])$ with $h_{\xi_0}'(1) = \frac{1-\gamma^2}{A-1}>0$.
\end{enumerate}
\end{lemma}

\proof 1.  The set $\Xi$ is not empty by Theorem \ref{ODE_existence}. The set $\Xi$  is bounded from above by Lemma \ref{Ricexplosion}.

\noindent 2. Let $h = h_\xi$ for some $\xi \in \Xi$. We argue by contradiction and assume there is $\hat y \in [0,1]$ such that
$$
h(\hat y) = \max_{y\in[0,1]} h(y) \ge 1.
$$
Because $h(0)=\gamma$ and $h(1) \le \gamma$, we have $\hat y \in (0,1)$ and so
$$
h(\hat y) > \gamma, \quad h'(\hat y) = 0,\quad h''(\hat y) \le 0.
$$
Inserting $h'(\hat y)=0$ into the ODE \eqref{wODE} produces the relation
$$
\frac{\xi}{\sigma_D^2} e^{\int_0^{\hat y} \frac{h(q)-1}{1-q}dq} =\frac{h(\hat y) \big(A\hat y h(\hat y)+1+\gamma -(2 \gamma +1) \hat y\big)-\gamma  (1+\gamma ) (1-\hat y)}{\hat y h(\hat y)^2}.
$$ 
Furthermore, inserting this expression into the second derivative of the ODE \eqref{wODE} yields
\begin{align*}
0&\ge \hat y^2 (1-\hat y)^2 h''(\hat y) \\
&= -\gamma  (1+\gamma ) (1-\hat y)^2 + \Big(1+\gamma +\hat y \big(\gamma  (3+\gamma) \hat y+\hat y-(1+\gamma) (2+\gamma)\big)\Big)h(\hat y) \\
&+\hat y \big(1+\gamma -\hat y (1+A+2 \gamma)\big)h(\hat y)^2 + A\hat  y^2h(\hat y)^3.
\end{align*}
First, because $A >1$ and $h(\hat y)\ge1$, 
$$
A\hat y^2\big( h(\hat y)^3 - h(\hat y)^2\big) \ge \hat y^2\big( h(\hat y)^3 - h(\hat y)^2\big).
$$
Therefore, 
\begin{align*}
0&\ge -\gamma  (1+\gamma) (1-\hat y)^2 + \Big(1+\gamma +\hat y \big(\gamma  (3+\gamma) \hat y+\hat y-(1+\gamma) (2+\gamma)\big)\Big)h(\hat y) \\
&+\hat y \big(\gamma +1-y (2 \gamma +2)\big)h(\hat y)^2 +\hat  y^2h(\hat y)^3\\
&=\big(h(\hat y) - \gamma\big)\Big(1+\gamma  + (1+\gamma) \big(h(\hat y)-2\big)\hat y +\big(h(\hat y)-1\big)\big(h(\hat y)-1-\gamma\big)\hat y^2\Big).
\end{align*}
Second, dividing by $\big(h(\hat y) - \gamma\big)>0$ implies the contradiction
\begin{align}\label{inequal17}
\begin{split}
0&\ge 1+\gamma  + (1+\gamma) \big(h(\hat y)-2\big)\hat y +\big(h(\hat y)-1\big)\big(h(\hat y)-1-\gamma\big)\hat y^2\\
&=(1+\gamma)(1-\hat y)  + (1+\gamma) \big(h(\hat y)-1\big)\hat y +\big(h(\hat y)-1\big)^2\hat y^2 - \gamma\big(h(\hat y)-1\big)\hat y^2\\
& > 0.
\end{split}
\end{align}
In \eqref{inequal17},  the strict inequality follows from 
$$
(1+\gamma)(1-\hat y) >0,\quad \big(h(\hat y)-1\big)^2\hat y^2 \ge0,
$$
and the sum of the remaining two terms in \eqref{inequal17} satisfies the bound
\begin{align*}
 \big(h(\hat y)-1\big)\big((1+\gamma)\hat y - \gamma\hat y^2\big)= \big(h(\hat y)-1\big)\big( \hat y +\gamma\hat y(1-\hat y)\big)\ge0.
\end{align*}

\noindent 3. We argue by contradiction and assume $\xi_0 \in \Xi$. 

\noindent{\bf Step 1/4:} Given $\epsilon \in (0,1-\gamma)$, there exists $y_0 \in (0,1)$ such that
$$
\forall y \in (y_0,1): \quad h_{\xi_0}(y) < \gamma + \epsilon<1.
$$
Because $h_{\xi_0}\le 1$, the function from \eqref{keyy2A}, i.e., 
$$
F_{\xi_0}(y) := \frac{\xi_0}{\sigma_D^2}e^{\int_0^y \frac{h_{\xi_0}(q)-1}{1-q}},\quad y \in [0,1),
$$
is non-increasing. Therefore, the limit $F_{\xi_0}(1):= \lim_{y\uparrow 1}F_{\xi_0}(y)$ exists and is zero because 
\begin{align*}
F_{\xi_0}(1)&= \lim_{y\uparrow 1}F_{\xi_0}(y_0)e^{\int_{y_0}^{y}\frac{h_{\xi_0}(q)-1}{1-q}dq}\\
&\le \lim_{y\uparrow 1} F_{\xi_0}(y_0)e^{\int_{y_0}^{y}\frac{\gamma +\epsilon-1}{1-q}dq}\\
&=  \lim_{y\uparrow 1}F_{\xi_0}(y_0)\left(  \frac{1-y_{0}}{1-y}\right)^{\gamma +\epsilon-1}\\
&=0.
\end{align*}
Because $A >1$, by increasing $y_0$ if necessary, we can therefore also assume $y_0$ satisfies
\begin{align}\label{keyy0}
\forall y \in (y_0,1): \quad F_{\xi_0}(y)< \frac{A-1}2.
\end{align}

 \noindent{\bf Step 2/4:} For $\xi\in (\xi_0,\xi_0+1)$, Theorem \ref{ODE_local_existence} gives a local solution $h_{\xi}(y)$ for $y \in [0,y_\xi]$ where $y_\xi$ is defined in \eqref{yxi}.  Lemma \ref{Lem_Lip} with $y_0\in(0,1)$ from the previous step and $\bar \xi := \xi_0+1$ gives a constant $M_1>0$ such that 
 \[
|h_{\xi}(y)-h_{\xi_0}(y)|\leq M_1 |\xi-\xi_0|,\quad y \le  y_{\xi} \land y_0,\quad \xi\in (\xi_0,\xi_0+1).
\]
Because $h_{\xi_0}(y) <1$ for all $y\in[0,1]$, we can consider $\xi \in (\xi_0,\xi_0+1)$ such that 
\begin{align}\label{keyy1}
 M_1 |\xi-\xi_0| + \sup_{y\in[0,1]} h_{\xi_0}(y) <1.
\end{align}
For such $\xi$ and $y \in   [0,y_\xi\land y_0]$, we have 
\begin{align*}
h_{\xi}(y) & \le |h_{\xi}(y) -h_{\xi_0}(y)| +h_{\xi_0}(y) \\
&\le M_1 |\xi- \xi_0| +h_{\xi_0}(y)\\
&<1.
\end{align*}
Therefore, whenever $\xi\in(\xi_0,\xi_0+1)$ satisfies \eqref{keyy1}, we have $y_\xi > y_0$.

 \noindent{\bf Step 3/4:} In addition to \eqref{keyy1}, this step creates an additional restriction on $\xi\in (\xi_0,\xi_0+1)$. First, let $F_\xi$ be defined as in \eqref{keyy2A}.  Lemma \ref{Lem_Lip} with $y_0\in(0,1)$ from the previous step and $\bar \xi := \xi_0+1$ gives a constant $M_2>0$ such that 
 \[
|F_{\xi}(y)-F_{\xi_0}(y)|\leq M_2 |\xi-\xi_0|,\quad y \le  y_{\xi} \land y_0,\quad \xi\in (\xi_0,\xi_0+1).
\]
For $\xi$ satisfying \eqref{keyy1},  the previous step ensures $y_{\xi} >y_0$ and so the bound in \eqref{keyy0} gives
\begin{align*}
F_{\xi}(y_0) &\le |F_{\xi}(y_0) -F_{\xi_0}(y_0) | + F_{\xi_0}(y_0) \\
&\le M_2 |\xi-\xi_0| +\frac{A-1}2.
\end{align*}
This shows that we can  lower $\xi\in (\xi_0,\xi_0+1)$ such that, in addition to \eqref{keyy1}, we also have 
\begin{align}\label{keyy2}
F_{\xi}(y_0)  <A -1.
\end{align}

 \noindent{\bf Step 4/4:} Finally, this step creates a contradiction using comparison of ODE solutions. Let $\xi\in (\xi_0,\xi_0+1)$ and $y_0\in (0,1)$ be as in the previous step. Using \eqref{keyy2}, the function in \eqref{keyy2A} satisfies
\begin{align*}
\forall y \in [y_0,y_\xi]:\quad F_\xi(y) = F_\xi(y_0) e^{\int_{y_0}^y \frac{h_\xi(q)-1}{1-q}}\le F_\xi(y_0)< A-1.
\end{align*} 
This bound gives the ODE inequality
\begin{align*}
h_\xi'(y) &= a_0(y) +  h_\xi(y)\frac{a_1(y) +\big(F_{\xi}(y)-A\big)h_\xi(y)}{1-y}\\
&\le a_0(y) +  h_\xi(y)\frac{a_1(y) -h_\xi(y)}{1-y},\quad y\in (y_0,y_\xi).
\end{align*} 
Using $a_3:=-1$ in  Theorem \ref{Thm:ODEbound} yields an ODE solution $f$ of
\begin{align*}
\begin{cases}
f'(y) = a_0(y) + f(y)  \frac{a_1(y)- f(y)}{1-y},\quad y\in (y_0,1),\\
f(y_0) = h_\xi(y_0).
\end{cases}
\end{align*}
The standard comparison principle gives $h_\xi(y) \le f(y)$ for $y\in  [y_0,y_\xi]$. Theorem \ref{Thm:ODEbound} and Remark \ref{rmkA}  give $f(1) =\gamma$ and $f(y) <1$ for $y \in [y_0,1]$ and so we have $y_\xi =1$. All in all, we have the contradiction $\xi> \xi_0$ and $\xi \in \Xi$ because
$$
h_{\xi}(1) \le f(1) = \gamma.
$$

\noindent{4.} Let $\xi_n \uparrow \xi_0:= \sup \Xi$ and let $h_{\xi_n}$ denote the corresponding solution of the ODE  \eqref{wODE}. Because of $h_\xi$'s monotonicity in $\xi$, we can define 
$$
h_{\xi_0}(y):= \lim_{n\to\infty} h_{\xi_n}(y),\quad y\in [0,1),
$$
which satisfies $0\le h_{\xi_0}(y) \le 1$ for $y\in [0,1)$. By taking limits in the ODE \eqref{wODE}, we see that the ODE
$$
h'_{\xi_0}(y) = a_0(y) + h_{\xi_0}(y)\frac{a_1(y) + \big(F_{\xi_0}(y)-A\big)h_{\xi_0}(y)}{1-y},\quad y \in (0,1),
$$
holds. We can argue as in first part of Step 2 in the proof of Theorem \ref{ODE_existence} to show that  $h_{\xi_0}(y)$ cannot oscillate as $y\uparrow1$ and so $h_{\xi_0} (1):=  \lim_{y\uparrow 1}h_{\xi_0}(y)$ exists.   Because $\xi_0 \not\in \Xi$, it must be that $h_{\xi_0} (1) >\gamma$.

To see  $h_{\xi_0} (1) =1$, the bound $h_{\xi_0} \le 1$ allows us to argue as in the second part of Step 2 in the proof of Theorem \ref{ODE_existence}. Here, it is shown that $h_{\xi_0} (1) =\ell_1$ where $\ell_1$ is defined in \eqref{ell}. The argument following \eqref{l1bound1} shows that $\ell_1<1$ implies the contradiction 
$$
h_{\xi_0} (1) = c_0 < \gamma <  h_{\xi_0} (1),
$$
where $c_0$ is defined in \eqref{c0}. Because $h_{\xi_0}(1) \le 1$, the only alternative is $\ell_1= 1$.

\noindent{5.}  This proof is similar to Step 2 in the proof of Theorem \ref{ODE_existence}. Because $0 \le h_{\xi_0} \le 1$, the limit $ \lim_{y \uparrow 1}e^{\int_0^y \frac{h_{\xi_0}(q)-1}{1-q}dq}$ exists in $[0,1]$, which allows us to use L'Hospital's rule below. The integral representation \eqref{z1a} of $h_{\xi_0}$ yields
$$
\frac{1}{y^{1+\gamma
}}\int_{0}^{y}x^{1+\gamma}(1-x)^{\gamma}\left(  a_{0}%
(x)+\frac{a_{2}(h_{\xi_0},x)}{1-x}h_\xi(x)^2\right)  \,dx= h_{\xi_0}(y)(1-y)^{\gamma}. 
$$
Because $h_{\xi_0}(y)$ is continuous at $y=1$ with a finite limit, the right-hand-side converges to 0 as $y\uparrow 1$. L'Hospital's rule give us
\begin{align*}
h_{\xi_0}(1) &=\lim_{y\uparrow1}\frac{1}{y^{1+\gamma
}(1-y)^{\gamma}}\int_{0}^{y}x^{1+\gamma}(1-x)^{\gamma}\left(  a_{0}%
(x)+\frac{a_{2}(h_{\xi_0},x)}{1-x}h_{\xi_0}(x)^2\right)  \,dx\\
& =\lim_{y\uparrow1}\frac{y^{1+\gamma}(1-y)^{\gamma}\left(  a_{0}%
(y)+\frac{a_{2}(h_{\xi_0},y)}{1-y}h_{\xi_0}(y)^2\right)  }{(1+\gamma)y^{\gamma}(1-y)^{\gamma
}-\gamma y^{1+\gamma}(1-y)^{\gamma-1}}\\
& =\lim_{y\uparrow1}\frac{y\big(  (1-y)a_{0}(y)+a_{2}(h_{\xi_0},y)h_{\xi_0}(y)^2\big)
}{(1+\gamma)(1-y)-\gamma y}\\
&=\frac1\gamma\left(A-\frac{\xi_0}{\sigma_D^2}\lim_{y \uparrow 1}e^{\int_0^y \frac{h_{\xi_0}(q)-1}{1-q}dq} \right)h_{\xi_0}(1)^2.
\end{align*}
Because $h_{\xi_0}(1)=1$, this implies \eqref{L0}.

The bounds $0\le h_{\xi_0}\le 1$ give
\begin{align}\label{hhhbound}
0\le e^{\int_{0}^{1}\frac{h(x)-1}{1-x}\,dx}\le1.
\end{align}
Because $h_{\xi_0}(0) =\gamma \in (0,1)$ and $h_{\xi_0}(y)$ is continuous at $y=0$, the
two inequalities in \eqref{hhhbound} are strict.

\noindent{6.}  Because $h_{\xi_0} \le 1$, the limit in \eqref{L0} gives the monotonicity property
\begin{align}\label{monotonicity_prop1}
e^{\int_0^y \frac{h_{\xi_0}(q)-1}{1-q}dq}\ge e^{\int_0^1 \frac{h_{\xi_0}(q)-1}{1-q}dq} = \frac{\sigma_D^2(A-\gamma)}{\xi_0},\quad y\in [0,1].
\end{align}
Therefore,
\begin{align*}
a_2(h_{\xi_0},y) &= \frac{\xi_0}{\sigma_D^2}e^{\int_0^y \frac{h_{\xi_0}(q)-1}{1-q}dq} - A\ge-\gamma.
\end{align*}
Using $a_3:=-\gamma$ in  Theorem \ref{Thm:ODEbound} produces an ODE solution $f$ of
\begin{align*}
\begin{cases}
f'(y) = a_0(y) +  f(y)\frac{a_1(y) -\gamma f(y)}{1-y},\quad y\in (0,1),\\
f(0) = \gamma,
\end{cases}
\end{align*}
Because $1+a_3 >0$, Theorem \ref{Thm:ODEbound} also gives $f$ with $f\ge \gamma$. The  comparison result in Theorem \ref{Thm:Comparison} gives the lower bound $h_{\xi_0}\ge f \ge \gamma$.

\noindent 7. We define
$$
g(y) := \frac{1-h_{\xi_0}(y)}{1-y},\quad y \in (0,1).
$$ 
\noindent{\bf Step 1/4:} To get an a priori estimate, this step assumes $h_{\xi_0}'(y)$ is continuous at $y=1$. The ODE in \eqref{wODE} and \eqref{L0} imply the representation
\begin{align}\label{C1_00}
\begin{split}
h_{\xi_0}'(y) &=  \frac{1+\gamma}y \big(\gamma - h_{\xi_0}(y) \big) +h_{\xi_0}(y)^2 (A-\gamma) \frac{e^{\int_y^1 g(q)dq} -1}{1-y}+\gamma h_{\xi_0}(y) g(y).
\end{split}
\end{align}
Because $h_{\xi_0}'(y)$ is continuous at $y=1$,  L'Hopital's rule gives
\begin{align*}
g(1) &= (1+\gamma)(\gamma-1)+(A-\gamma)\lim_{y\uparrow 1} \frac{e^{\int_y^1 g(q)dq }-1}{1-y} +\gamma g(1)\\
&=\gamma^2-1 + (A-\gamma)g(1) + \gamma g(1).
\end{align*}
We can solve to see  $g(1) = \frac{1-\gamma^2}{A-1}>0$. The next steps prove that $h_{\xi_0}'(y)$ is indeed continuous at $y=1$.

\noindent{\bf Step 2/4:} This step rules out that $g(y)$ can monotonically explode to $\infty$ as $y\uparrow 1$. We argue by contradiction. The Mean-Value Theorem and $g'(y)>0$ for $y$ near 1 yields $\theta(y) \in (y,1)$ such that
\begin{align*}
e^{\int_y^1 g(q)dq} -1&= e^{\int_y^1 g(q) dq}g\big(\theta(y)\big)(1-y) \ge e^{\int_y^1g(q)dq}g(y)(1-y),
\end{align*}
 for $y$ near 1. The ODE in \eqref{C1_00} gives
\begin{align}\label{C1_B1}
\begin{split}
h_{\xi_0}'(y) &\ge  \frac{1+\gamma}y \big(\gamma - h_{\xi_0}(y) \big) +\Big( h_{\xi_0}(y)^2(A-\gamma)e^{\int_y^1 g(q)dq} +\gamma h_{\xi_0}(y)\Big)g(y),
\end{split}
\end{align}
for $y$ near 1. By differentiation, we see 
\begin{align}\label{gprime34}
g'(y) = \frac{1}{1-y}\big( g(y) - h'_{\xi_0}(y)\big)
\end{align} 
and so $g'(y) >0$ implies that  $g(y)\ge  h'_{\xi_0}(y)$ for $y$ near 1. The inequality \eqref{C1_B1} gives
\begin{align}\label{C1_B1q}
g(y) &\ge  \frac{1+\gamma}y \big(\gamma - h_{\xi_0}(y) \big) +\Big( h_{\xi_0}(y)^2(A-\gamma)e^{\int_y^1 g(q)dq} +\gamma h_{\xi_0}(y)\Big)g(y),
\end{align}
for $y$ near 1. Because $g\ge0$, we have $e^{\int_y^1 g(q)dq}\ge 1$ and so the inequality \eqref{C1_B1q} gives
\begin{align*}
&\frac{1+\gamma}y \big(h_{\xi_0}(y)  -\gamma \big) \ge  \big( h_{\xi_0}(y)^2(A-\gamma) +\gamma  h_{\xi_0}(y)-1\big)g(y),
\end{align*}
for $y$ near 1. By taking limits as $y\uparrow 1$ and using $A>1$ give the contradiction
\begin{align*}
(1+\gamma)(1-\gamma) &= \lim_{y\uparrow 1}\frac{1+\gamma}y \Big(h_{\xi_0}(y)  -\gamma \Big) \ge ( A -1)\lim_{y\uparrow 1}g(y) = \infty.
\end{align*}

\noindent{\bf Step 3/4:} This step considers  oscillations. The ODE \eqref{C1_00} and \eqref{gprime34} give
\begin{align}\label{gprime55}
\begin{split}
- (1-y)g'(y)&= \frac{1+\gamma}y \big(\gamma - h_{\xi_0}(y) \big) +h_{\xi_0}(y)^2 (A-\gamma) \frac{e^{\int_y^1 g(q)dq} -1}{1-y}\\
&+\big(\gamma h_{\xi_0}(y) -1\big)g(y),\quad y\in(0,1).
\end{split}
\end{align}
Differentiating \eqref{gprime55} at a point $y\in(0,1)$ with $g'(y) = 0$ gives
\begin{align}\label{gprime35}
\begin{split}
(1-y)^2yg''(y)&=(1+\gamma )\big( \gamma - h_{\xi_0}(y)\big) -h_{\xi_0}(y)^2(A-\gamma ) \left(e^{\int_y^1 g(q)dq}-1\right) \\
&+ \bigg( h_{\xi_0}(y) \left(y (A-\gamma ) e^{\int_y^1 g(q)dq} \big(h_{\xi_0}(y)-2\big)+2 A y-\gamma \right)\\
&+\gamma -(\gamma +3) y+2- \gamma  (1-y) y g(y)\bigg)g(y) ,\quad y\in(0,1).
\end{split}
\end{align}

First, assume  that $(y_n)_{n\in\N}$ is a sequence of local maxima for $g$ with  $y_n \uparrow 1$ such that
$$
 \forall n\in\N: \quad g'(y_n) =0,\quad g''(y_n) \le 0.
$$
We can extract a subsequence $(y_{n_k})_{k\in\N} \subseteq (y_n)_{n\in\N}$ such that
$$
\lim_{k\to\infty} g(y_{n_k}) =  \limsup_{n\to\infty} g(y_n) \in [0,\infty].
$$ 
Because $h_{\xi_0}(1) =1$, we have
$$
\lim_{k\to\infty} (1-y_{n_k}) y_{n_k} g(y_{n_k}) = \lim_{k\to\infty}  y_{n_k}\big( 1- h_{\xi_0}(y_{n_k})\big) = 0,
$$
and so replacing $y$ with $y_{n_k}$ in \eqref{gprime35} and taking limits give
\begin{align*}
0 &\ge \lim_{k\to\infty}  (1-y_{n_k})^2y_{n_k}g''(y_{n_k}) =\gamma^2-1 + (A-1) \lim_{k\to\infty} g(y_{n_k}).
\end{align*}
This gives the upper bound for $g$'s local maxima
\begin{align}\label{bound1}
\limsup_{n\to\infty} g(y_n) \le \frac{1-\gamma^2}{A-1}.
\end{align}

Second, assume that $(y_n)_{n\in\N}$ is a sequence of local minima for $g$ with  $y_n \uparrow 1$ such that
$$
 \forall n\in\N:\quad  g'(y_n) =0,\quad g''(y_n) \ge 0.
$$
Then, similarly to \eqref{bound1}, we have the lower bound for $g$'s local minima
\begin{align}\label{bound2}
\liminf_{n\to\infty}  g(y_n) \ge \frac{1-\gamma^2}{A-1}.
\end{align}

Finally, combining the two bounds  \eqref{bound1} and \eqref{bound2} shows that should $g(y)$ oscillate as $y\uparrow 1$, the function $g$ will still have limit $\lim_{y\uparrow 1}g(y) = g(1) = \frac{1-\gamma^2}{A-1}$ as in  Step 1.

\noindent{\bf Step 4/4:} This step shows that $h_{\xi_0}(y)$ is continuously differentiable at $y=1$. Step 2 and Step 3 ensure that that $h'_{\xi_0}(1)$ exists and that $g(y)$ is continuous at $y=1$. It is $g$'s continuity which allows us to use L'Hopital's rule  to see
\begin{align*}
\lim_{y\uparrow 1} \frac{\gamma + \big((A-\gamma)e^{\int_y^1 g(q)dq }-A\big)h_{\xi_0}(y)}{1-y}
&= 
\lim_{y\uparrow 1} \frac{ (A-\gamma)\big(e^{\int_y^1 g(q)dq }-1\big)h_{\xi_0}(y) + \gamma\big(1-h_{\xi_0}(y) \big)}{1-y}\\
&=(A-\gamma)\lim_{y\uparrow 1} \frac{e^{\int_y^1 g(q)dq }-1}{1-y} +\gamma \lim_{y\uparrow 1} \frac{1-h_{\xi_0}(y)}{1-y}\\
&=(A-\gamma)g(1) + \gamma g(1)\\
&= Ag(1).
\end{align*}
Then, the ODE  \eqref{C1_00} and $h_{\xi_0}(1)=1$ give
\begin{align}\label{C1_E}
\lim_{y\uparrow 1} h_{\xi_0}'(y) = (1+\gamma)(\gamma-1) + Ag(1)= \gamma^2-1 + Ag(1).
\end{align}
Another application of L'Hopital's rule gives
\begin{align}\label{C1_E2}
g(1) = \lim_{y\uparrow 1}\frac{1-h_{\xi_0}(y)}{1-y} = \lim_{y\uparrow 1} h_{\xi_0}'(y) = \gamma^2-1 + Ag(1).
\end{align}
By solving \eqref{C1_E2}, we get $g(1) = \frac{1-\gamma^2}{A-1}>0$ and so \eqref{C1_E} gives $\lim_{y\uparrow 1} h_{\xi_0}'(y) = \frac{1-\gamma^2}{A-1}$.
$\endproof$

\newpage
Figure 1 presents a numerical illustration.
\begin{figure}[!h]\label{Fig_h}
\begin{center}
\caption{ODE solutions $h_{\xi_n}$ for $\xi_n \uparrow \sup \Xi$. The parameters are $\gamma:=0.5, \sigma_D:=0.2, A:=2$, and $\xi_n \in \{0.15, 0.152, 0.1522, 0.15223, 0.152232\}$.}
\ \\
$\begin{array}{c}
\includegraphics[width=6cm, height=4.5cm]{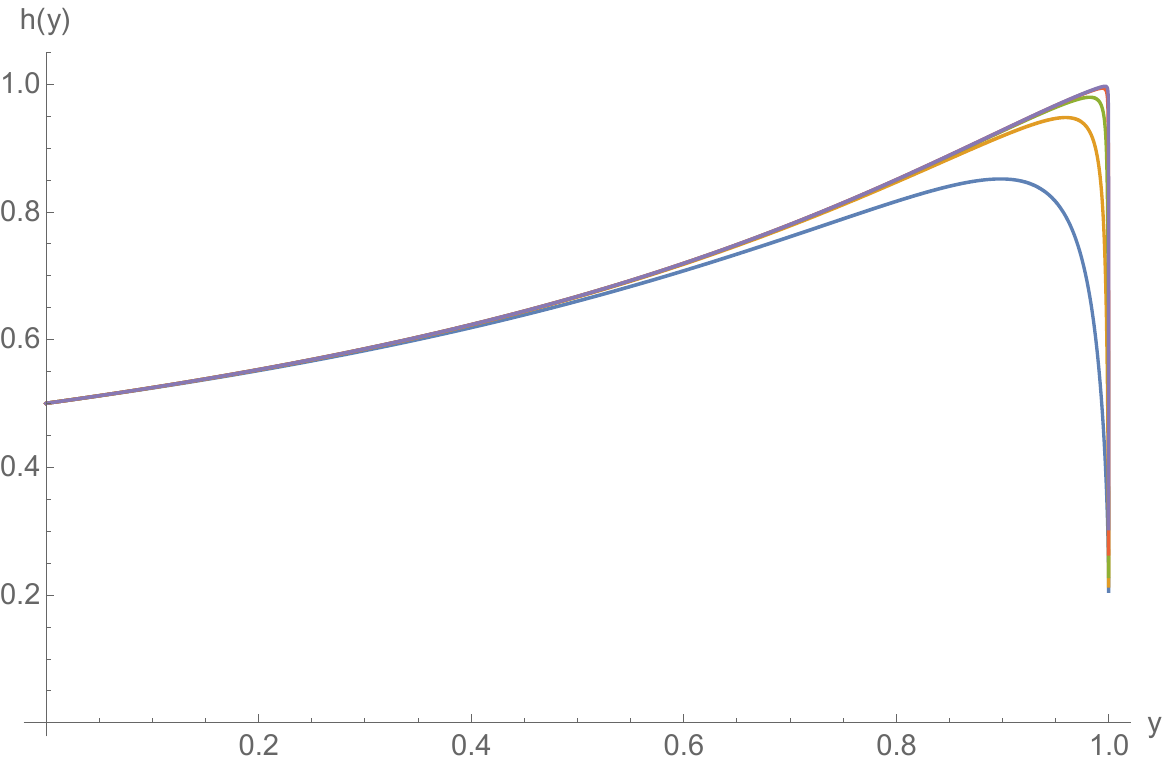} 
 \end{array}$
\end{center}
\end{figure}
\vspace{-1.5cm}


\section{Application to incomplete Radner equilibrium theory}\label{sec:application}
On a probability space $(\Omega,\mathbb{F},\P)$, we let $(B_t)_{t\ge0}$ be a Brownian motion generating the raw filtration $(\sF^0_t)_{t\ge0}$ with $\sF_t^0 := \sigma(B_s)_{s\in [0,t]}$ and assume  $\mathbb{F} = \vee_{t\ge0} \sF_t^0$. We  define the augmented filtration as $\sF_t:=\sF_t^0\vee\mathbb{F}$'s $\P$-nullsets for $t\ge0$. For $t\ge0$, we write $\E_t[\cdot]$ as short-hand notation for the conditional expectation $\E[\cdot |\sF_t]$.

Throughout this section, let $h$ be the solution in Theorem \ref{theorem_main_h}.

\subsection{Autonomous state process $Y$}
We define the drift function $\mu_Y$ and and volatility function $\sigma_Y$ as 
\begin{align}\label{mgmethod1bbq}
\begin{split}
\mu_Y(y)&:=\frac{\sigma_D^2 (1-y) \big(1+\gamma +2 \gamma  y h(y)-2 y(1+\gamma)\big)}{2 y h(y)^2},\quad y\in(0,1],\\
\sigma_Y(y)&:= \sigma_D\frac{1-y}{h(y)},\quad y\in[0,1].
\end{split}
\end{align}
The next subsections use an autonomous state process $Y = (Y_t)_{t\ge0}$ valued in $[0,1]$ with dynamics
\begin{align}\label{inhomeY}
dY_t = \mu_Y(Y_t)dt +\sigma_Y(Y_t) dB_t,\quad Y_0 \in(0,1).
\end{align} 
It is $\mu_Y(y)$'s singularity at $y=0$ and the singularities of both $\mu_Y(y)$ and $\sigma_Y(y)$ at $y=1$ that are problematic in the following analysis. The boundary behavior of the ratio 
 \begin{align}\label{singula0}
\frac{\mu_Y(y)}{\sigma_Y(y)^2}=\frac{1+\gamma +2 \gamma  y h(y)-2 y(1+\gamma)}{2 y(1-y)},\quad y\in(0,1),
\end{align}
as $y\downarrow 0$ and $y\uparrow1$ classifies $Y$'s boundary behavior. We have the limits
\begin{align}\label{singula1}
\begin{split}
\lim_{y\downarrow 0}y\frac{\mu_Y(y)}{\sigma_Y(y)^2}&=\frac{1+\gamma}{2}>0,\\
\lim_{y\uparrow 1}(1-y)\frac{\mu_Y(y)}{\sigma_Y(y)^2}&=-\frac{1}{2}(1-\gamma)<0.
\end{split}
\end{align}
Example 5.4 in \cite{Cherny_Engelbert} illustrates that the singularities $\frac1y$ as $y\downarrow 0$ and $\frac1{1-y}$ as $y\uparrow 1$  in \eqref{singula1} require a  careful error analysis. In particular, only identifying the limits in \eqref{singula1} is insufficient to classify the boundary points $y\in \{0,1\}$ for the SDE \eqref{inhomeY}. The next subsections establish the following result.

\begin{theorem}\label{strong_lemma} Let $\gamma \in (0,1)$, $\sigma_D^2>0$, and $A>1$.  For any initial value $Y_0 \in [0,1)$, there exists a  strong solution $Y$, pathwise unique, of the SDE \eqref{inhomeY} for the coefficients defined in  \eqref{mgmethod1bbq}.
\end{theorem}

\subsubsection{Left-boundary point for $Y$}
We start with the left-boundary point $y=0$.

\begin{lemma}\label{lem_tech1}  Let $\gamma \in (0,1)$, $\sigma_D^2>0$, and $A>1$. The functions in \eqref{mgmethod1bbq}  satisfy 
\begin{align}\label{expansion}
y\frac{\mu_Y(y)}{\sigma_Y(y)^2} &= b_0 + b_1(y)y,\quad 
b_0 := \frac{1+\gamma}2,\quad b_1(y) := \frac{2 \gamma  h(y)-1-\gamma }{2 (1-y)},\quad y\in(0,1).
\end{align}
The boundary point $y=0$ is inaccessible and  entrance. Furthermore,  the SDE \eqref{inhomeY} has a nonnegative weak solution, unique in law, starting from $Y_0=0$.

\end{lemma}

\proof Because $b_1$ is uniformly lower bounded on any interval $[0,a]$ for $a\in (0,1)$, we can define a constant $M\in\R$ as
$$
M:=\inf_{y\in[0,a]} b_1(y).
$$
As in Eq. (2.12) in \cite{Cherny_Engelbert}, we define the function 
\begin{align}\label{rho}
\rho(y) :=\exp\left(2\int_y^a \frac{\mu_Y(z)}{\sigma_Y(z)^2} dz\right), \quad y\in(0,a].
\end{align}
The decomposition in \eqref{expansion} gives
\begin{align}\label{rholow1}
\begin{split}
\rho(y) &=\exp\left(2\int_y^a \frac{\mu_Y(z)}{\sigma_Y(z)^2} dz\right)\\
&=\exp\left(2\int_y^a\Big( \frac{b_0 }z+ b_1(z)\Big) dz\right)\\
&\ge  \exp\left(2\int_y^a \Big(\frac{b_0}z +M\Big) dz\right)\\
&=  \exp\bigg(2b_0 \Big(\log(a)-\log(y)\Big)\bigg)\exp\big(2M(a-y)\big)\\
&=  \exp\bigg(\log\Big(\big(\frac{a}{y}\big)^{2b_0 }\Big)\bigg)\exp\big(2M(a-y)\big)\\
&= \Big(\frac{a}{y}\Big)^{2b_0 }\exp\big(2M(a-y)\big).
\end{split}
\end{align}
Following Eq. (2.13) in  \cite{Cherny_Engelbert}, we define the scale function
\begin{align}\label{scale_fct1}
s(y) := - \int_y^a \rho(z) dz,\quad y\in (0,a].
\end{align}
Because $b_0 \ge \frac12$, we get $s(0):=\lim_{y\downarrow 0} s(y) = -\infty$. Therefore, Theorem 8A in \cite{Helland1996} ensures that $y=0$ is inaccessible.

Next, we first calculate the derivative
\begin{align*}
\rho'(y) &=\frac{\partial}{\partial y}\exp\left(2\int_y^a \Big(\frac{b_0}z + b_1(z)\Big) dz\right)\\
&=-2\rho(y) \Big(\frac{b_0}y + b_1(y)\Big).
\end{align*}
The lower bound \eqref{rholow1} gives
\begin{align}\label{rholow2}
y\rho(y) \ge  y\big(\frac{a}{y}\big)^{2b_0 }\exp\big(2M(a-y)\big).
\end{align}
Because $b_0 > \frac12$, the lower bound \eqref{rholow2} gives $\lim_{y\downarrow 0} y\rho(y) = \infty$.

Second, two applications of  L'Hopital's rule yield
\begin{align*}
\lim_{y\downarrow 0}\frac{ |s(y)|}{\rho(y)} &= \lim_{y\downarrow 0} \frac{ 1}{\rho(y)}\int_y^a \rho(z) dz \\
&=- \lim_{y\downarrow 0} \frac{ 1}{\rho'(y)} \rho(y) \\
&=\lim_{y\downarrow 0} \frac{ y}{2\big(b_0 + b_1(y)y\big)}\\
&=0,\\
\lim_{y\downarrow 0}\frac{ |s(y)|}{y\rho(y)} &= \lim_{y\downarrow 0} \frac{ 1}{y\rho(y)}\int_y^a \rho(z) dz \\
&=- \lim_{y\downarrow 0} \frac{ 1}{\rho(y)+ y\rho'(y)} \rho(y) \\
&=- \lim_{y\downarrow 0} \frac{ 1}{1-2y \Big(\frac{b_0}y + b_1(y)\Big)} \\
&=\frac{ 1}{2b_0-1} \in (0,\infty),
\end{align*}
where the last line follows from $b_0 > \frac12$. Because $b_1$ is bounded and $h\ge\gamma$, we have $\sigma_Y(y) \ge \frac{\sigma_D(1-y)}{\gamma}$ and
\begin{align*}
\int_0^a \frac{1+|\mu_Y(y)|}{\rho(y)\sigma_Y(y)^2} |s(y)|dy  &\le\int_0^a \bigg(\frac{1}{\rho(y)\sigma_Y(y) ^2}  + \frac{\frac{b_0}{y} + |b_1(y)|}{\rho(y)}\bigg) |s(y)|dy <\infty.
\end{align*}
Theorem 8C in \cite{Helland1996} gives us that $y=0$ is not natural (equivalently, $y=0$ is an entrance point) because 
$$
\int_0^a \frac{1}{\rho(y)\sigma_Y(y)^2} |s(y)|dy  <\infty.
 $$
 
 Finally,  Theorem 2.16 in \cite{Cherny_Engelbert} shows that $Y_0=0$ produces a nonnegative weak solution, unique in law, of the SDE \eqref{inhomeY}.

$\endproof$

\subsubsection{Right-boundary point for $Y$}

\begin{lemma}\label{lem2}  Let $\gamma \in (0,1)$, $\sigma_D^2>0$, and $A>1$.  Then, the boundary point $y=1$ is inaccessible, natural, and attracting.
\end{lemma}

\proof We let $a\in (0,1)$ be arbitrary and consider the function\footnote{The definition of $\rho$ in \eqref{rholow3} coincides with the definition of $\rho$ in \eqref{rho}.}
\begin{align}\label{rholow3}
\begin{split}
\rho(x) :=&\exp\left(-2\int_a^x \tfrac{\mu_Y(y)}{\sigma_Y(y)^2} dy\right)\\
=&\exp\left(\int_a^x \Big(\tfrac{(2 y-1)(1+\gamma)-2 \gamma  y h(y)}{(1-y) y}\Big) dy\right)\\
=&\exp\left(\int_a^x \Big(\tfrac{(2 y-1)(1+\gamma)-2\gamma y -2 \gamma y \big(h(y)-1\big)}{(1-y)y}\Big) dy\right)\\
=&\exp\left(\int_a^x \Big(\tfrac{2-\frac1y(1+\gamma) -2 \gamma \big(h(y)-1\big)}{1-y}\Big) dy\right),\quad  x\in [a,1).
\end{split}
\end{align}
Taylor's formula for $\frac1y$ centered at $y=1$ produces $
\frac1y = 1 + (1-y) + o(1-y)$. 
Therefore, 
$$
\int_a^x \frac1{(1-y)y}dy = \int_a^x \left(\frac1{1-y} + 1 + \frac{o(1-y)}{1-y}\right)dy,\quad x\in [a,1].
$$
This allows us to rewrite \eqref{rholow3} as
\begin{align}\label{rholow37}
\begin{split}
\rho(x)&=\exp\left(\int_a^x \Big(\tfrac{1-\gamma -2 \gamma\big(h(y)-1\big)}{1-y}\Big) dy - (1+\gamma)\Big( x-a  + \int_a^x \tfrac{o(1-y)}{1-y}dy\Big)\right)\\
&= \exp\left(\int_a^x \tfrac{1-\gamma }{1-y} dy+2\gamma\int_a^x \tfrac{1-h(y)}{1-y} dy - (1+\gamma)\Big( x-a  + \int_a^x \tfrac{o(1-y)}{1-y}dy\Big)\right)\\
&= \Big(\frac{1-a}{1-x}\Big)^{1-\gamma} \exp\big(C(x)\big),
\end{split}
\end{align}
where we have defined the function
\begin{align*}
C(x) := 2\gamma\int_a^x \tfrac{1-h(y)}{1-y} dy - (1+\gamma)\Big( x-a + \int_a^x \tfrac{o(1-y)}{1-y}dy\Big),\quad x\in [a,1].
\end{align*}
Lemma \ref{lemma:xi0}(5) ensures that $C(x)$ is a continuous function for $x\in [a,1]$ and so 
\begin{align}\label{rholow38}
c_1 := \min_{x\in [a,1]}C(x)\in \R,\quad  c_2 := \max_{x\in [a,1]} C(x)\in \R.
\end{align}
Because $\gamma \in (0,1)$, we see from \eqref{rholow37} that $\rho(x)$ is integrable at $x=1$ and so the function\footnote{The definition of $s$ in \eqref{scalefct} coincides with the definition of $s$ in \eqref{scale_fct1}.}
\begin{align}\label{scalefct}
s(y) := \int_a^y \rho(x) dx,\quad y\in [a,1),
\end{align}
satisfies $s(1):= \lim_{y\uparrow 1}s(y) <\infty$.

Because Lemma \ref{lemma:xi0}(6) ensures $h\ge \gamma$, we have 
$$
\sigma_Y(y)^2 = \frac{\sigma_D^2 (1-y)^2}{h(y)^2} \le \frac{\sigma_D^2(1-y)^2 }{\gamma^2},\quad y\in [0,1].
$$
The Mean-Value Theorem gives $\theta(y) \in [y,1]$ such that
\begin{align*}
s(1) - s(y) &= s'\big(\theta(y)\big)(1-y) \\
&= \rho\big(\theta(y)\big) (1-y) \\
&= \Big(\frac{1-a}{1-\theta(y)}\Big)^{1-\gamma} e^{C\big(\theta(y)\big)} (1-y)\\
&\ge  \Big(\frac{1-a}{1-\theta(y)}\Big)^{1-\gamma} e^{c_1}(1-y).
\end{align*}
For some constant $c_3>0$, by \eqref{rholow37}, \eqref{rholow38}, and $h(1)=1$, we have
\begin{align}\label{scalefctV}
\begin{split}
\frac{s(1) - s(y)}{\rho(y)\sigma_Y(y)^2 } &\ge \frac{c_3(1-y)}{\big(1-\theta(y)\big)^{1-\gamma}(1-y)^{1+\gamma} } \\
 &\ge  \frac{c_3(1-y)}{(1-y)^{1-\gamma}(1-y)^{1+\gamma} }\\
 &= \frac{c_3}{1-y },
 \end{split}
\end{align}
which is not integrable for $y$ near 1. Therefore, Theorem 8A in \cite{Helland1996} gives that the boundary point $y=1$ is inaccessible.

From \eqref{rholow37}, we $\rho(x) \ge c_4 (1-x)^{\gamma-1}$ for some constant $c_4>0$. This gives the lower bound 
\begin{align*}
\frac{s(y)}{\rho(y)\sigma_Y(y)^2}  &\ge c_5 \frac{\int_a^y  (1-x)^{\gamma-1}dx}{(1-y)^{1+\gamma}} =  c_5 \frac{(1-a)^{\gamma}-(1-y)^{\gamma}}{\gamma(1-y)^{1+\gamma}}
\end{align*}
for some constant $c_5$. Because this lower bound is not integrable for $y$ near 1, Theorem 8C in \cite{Helland1996} gives that $y=1$ is natural (i.e., $y=1$ is not entrance).

Finally, Theorem 8D in \cite{Helland1996} gives that $y=1$ is attracting because $s(1)<\infty$ and the lower bound
\begin{align*}
\frac{1}{\rho(y)\sigma_Y(y)^2}  &\ge c_6 \frac{1}{(1-y)^{1+\gamma}}
\end{align*}
is not integrable for $y$ near 1, where $c_6$ is another constant.

$\endproof$

\subsubsection{Proof of strong SDE existence of $Y$}

The proof of the next result uses a variation of Le Gall's uniqueness argument based on local times (see, e.g., Exercise 3.7.12 in \cite{KS1988}). 

\proof[Proof of Theorem \ref{strong_lemma}]

 For $Y_0 \in (0,1)$, the Engelbert and Schmidt conditions (i.e., non-degeneracy and local integrability) hold and ensure that a weak solution exists, unique in law,  see, e.g., Theorem 5.5 in \cite{KS1988}. For $Y_0=0$, there is also a weak solution, unique in law,  by Lemma \ref{lem_tech1}.

To upgrade these weak solutions to strong solutions it suffices to prove pathwise uniqueness locally. To this end, for $n\in \N$ such that $Y_0 \in [0, 1-\frac1n]$, we define the stopping times $\tau_n := \inf \{ t\ge0 :  Y_t =1-\frac1n\}$ and will prove that strong uniqueness holds for $t\in [0, \tau_n]$. Because $h,h' \in \sC([0,1))$ with $h\ge \gamma$, the formula
$$
\sigma_Y'(y) =-\sigma_D \frac{ (1-y) h'(y)+ h(y)}{h(y)^2},\quad y\in [0,1-\tfrac1n],
$$
shows that $\sigma_Y(y)$ is Lipschitz for $y\in [0, 1-\frac1n]$.  Pathwise uniqueness follows from  Proposition  IX.3.2 in \cite{Revuz_Yor} because the local-time condition $L^0(Y^1-Y^2)=0$ holds by  $\sigma_Y$'s Lipschitz property and Corollary  IX.3.4 in \cite{Revuz_Yor}.

$\endproof$

\subsection{Investors' maximization problems}

There is a single consumption good that serves as our model's num\'eraire (i.e., all prices are real).  As in \cite{BC1998}, the stock pays dividends at rate $D =(D_t)_{t\ge0}$ given by the geometric Brownian motion 
\begin{align}\label{dD}
dD_t := D_t \big(\mu_D dt + \sigma_D dB_t\big),\quad D_0>0,
\end{align}
where $D_0>0$, $\mu_D\in\R$, and $\sigma_D>0$ are exogenous constants.\footnote{As in \cite{BC1998}, the investors do not receive personal income and so $D$ is the model's aggregate consumption-rate process.}

There are two assets: a money market account with price process $S^{(0)}=(S^{(0)})_{t\ge0}$ and a stock with price process $S=(S_t)_{t\ge0}$ paying dividends at rate $D$. We conjecture (and verify) that there exist It\^o processes $(S^{(0)},S)$ with dynamics
\begin{align}
dS^{(0)}_t &= r_t S^{(0)}_t dt, \quad S_0^{(0)}:=1,\label{dS0}\\
dS_t &= (S_t r_t -D_t)dt +S_t \sigma_{S,t} (\kappa_{t}dt+dB_t),\quad S_0\in (0,\infty),\label{dS}
\end{align}
where $S_0$ is a constant and $(r,\kappa,\sigma_{S})$ are stochastic processes. Below, $(S_0,r,\kappa,\sigma_{S})$  will be endogenously determined in equilibrium. To ensure that  \eqref{dS0} is well-defined, we require $r\in \sL^1_\text{loc}$.\footnote{A process $Q = (Q_t)_{t\ge0}$ belongs to $\sL_\text{loc}^p$, $p\in \{1,2\}$, if $Q$ is progressively measurable and satisfies $\P(\int_0^T |Q_t|^p dt <\infty)=1$ for all $T>0$.} To ensure that \eqref{dS} is well-defined, we require $\sigma_S\in \sL^2_\text{loc}$ and  $\kappa \in  \sL^2_\text{loc}$.\footnote{Unlike the literature on endogenous market  completeness (see, e.g., \cite{Anderson-Raimondo}),  we do not require $\sigma_S\neq0$.}

As in \cite{BC1998}, we consider a model with only two investors (equivalently, two groups of homogenous investors), and both investors are price takers. Investor 1 can trade the stock whereas investor 2 cannot hold or trade the stock. Investor 1's self-financing wealth process is defined as
\begin{align}\label{dW1}
\begin{split}
dX_{1,t} & := r_t X_{1,t} dt +  \theta_{1,t}S_{t}  \sigma_{S,t}(\kappa_{t}dt+ dB_t) - c_{1,t}dt,\\ 
X_{1,0}&:= \theta^{(0)}_{1,0-} +S_0\in\R,
\end{split}
\end{align}
where the nonnegative consumption-rate process $c_1\in \sL^1_\text{loc}$ and the number of shares of stock  held $\theta_1$ are investor  1's control processes (both $c_1$ and $\theta_1$ are endogenously determined in equilibrium). In \eqref{dW1}, the exogenous constant $\theta^{(0)}_{1,0-} $ denotes investor 1's endowed holdings in the money market account. Furthermore, for simplicity, we normalize the stock supply to one share outstanding so that $X_{1,0}$ in \eqref{dW1} reflects investor 1's endowed stock holdings being $\theta_{1,0-}=1$.

The dynamics of investor  2's self-financing wealth process are simpler because she cannot access the stock market at any time. Investor  2's self-financing wealth process is defined by
\begin{align}\label{dW2}
\begin{split}
dX_{2,t} & := r_tX_{2,t} dt  - c_{2,t}dt,\\ 
X_{2,0}&:= \theta^{(0)}_{2,0-}\in(0,\infty).
\end{split}
\end{align}
In \eqref{dW2}, the nonnegative process $ c_{2}\in \sL^1_\text{loc}$ is investor  2's consumption-rate  process  and is endogenous,  and $\theta^{(0)}_{2,0-}$ is  investor  2's endowed holdings in the money market account (exogenous). 
The process $c_2$ is endogenously determined in equilibrium and must be such that the dynamics \eqref{dW2} are well-defined. Because the money market account is in zero-net supply, the investors' endowed money market holdings satisfy $\theta^{(0)}_{1,0-}+\theta^{(0)}_{2,0-}=0$. The restriction on the initial money market holdings $ \theta^{(0)}_{2,0-} = -  \theta^{(0)}_{1,0-}>0$ in \eqref{dW2} is taken from footnote 5 in  \cite{BC1998}.

As in  \cite{BC1998}, investor 1 and investor 2 are assumed to have identical preferences. However, unlike    the log-log utilities in \cite{BC1998}, our model uses a common time preference parameter $\beta>0$ and a common constant relative risk aversion coefficient $\gamma \in (0,1)$.\footnote{There are two natural extensions of our setting: (i) extend the model to different powers for the two investors, and (ii) extend the model to $\gamma\in (1,\infty)$. However, both such extensions produce more complicated governing ODEs than \eqref{KEY_ODE} and so we leave these extensions to future research.} The investors' maximization problems are
\begin{align}\label{primalval}
&\sup_{\theta_1,c_1\in \sA_1}\tfrac{1}{1-\gamma}\E\left[\int_0^\infty e^{-\beta t}c^{1-\gamma}_{1,t}dt\right],\quad \sup_{ c_2\in \sA_2}\tfrac{1}{1-\gamma}\E\left[\int_0^\infty e^{-\beta t}c^{1-\gamma}_{2,t}dt\right],
\end{align}
where the admissible sets $\sA_1$ and $\sA_2$ in \eqref{primalval} are defined as:

 \begin{definition}[Admissibility] \label{def:ad} We deem progressively measurable processes $(\theta_1,c_1)$ \emph{admissible} for investor 1 and we write $(\theta_1,c_1) \in \sA_1$ if $c_{1,t}\ge0$, $t\ge0$, and the solution of \eqref{dW1} is nonnegative, i.e., $X_{1,t}\ge0$ for all $t\ge0$. Likewise, we deem a progressively measurable process $c_2$ \emph{admissible}  for investor 2 and we write $c_2 \in \sA_2$ if  $c_{2,t}\ge0$, $t\ge0$, and the solution of \eqref{dW2} is nonnegative,  i.e., $X_{2,t}\ge0$ for all $t\ge0$. 
$\endproof$
\end{definition}

\subsection{Radner equilibrium}
Because both investors are price takers, the equilibrium is referred to as a Radner equilibrium. This equilibrium concept is standard and can be found in, e.g., Chapter 4 in the textbook \cite{KS1998}. 

We recall that the stock supply is in a net supply of one, the money market account is in zero net supply, and the aggregate consumption rate is $D$ in \eqref{dD}.
  
\begin{definition}[Radner equilibrium] \label{def:eq} A constant $S_0>0$  and processes $(r,\kappa,\sigma_{S})$ and $(\hat \theta_1, \hat c_1, \hat c_2)$ constitute a Radner equilibrium if:
\begin{itemize}
\item[(i)] The processes $(\hat{\theta}_1,\hat{c}_1)\in \sA_1$ maximize \eqref{primalval} for $i=1$; that is,
\begin{align}\label{primalval1}
(\hat{\theta}_1,\hat{c}_1)\in \text{argmax}_{(\theta,c)\in\sA_1}\tfrac{1}{1-\gamma} \E\Big[\int_0^\infty e^{-\beta t} c^{1-\gamma}_t dt\Big]<\infty.
\end{align}
\item[(ii)] The process $\hat{c}_2\in \sA_2$ maximizes \eqref{primalval} for $i=2$; that is,
\begin{align}\label{primalval2}
\hat{c}_2\in \text{argmax}_{c\in\sA_2} \tfrac{1}{1-\gamma}\E\Big[\int_0^\infty e^{-\beta t}c^{1-\gamma}_t dt\Big]<\infty.
\end{align}

\item[(iii)] The stock and consumption markets clear in the sense
\begin{align}
\hat{\theta}_{1,t} =1\quad \text{and }\quad \hat{c}_{1,t}+\hat{c}_{2,t}=D_t,\quad \text{for all }t\ge0.\label{clearingconds}
\end{align}
\end{itemize}
\vspace{-0.5cm}
$\endproof$
\end{definition}

\noindent{Walras' law ensures that clearing in the stock market and clearing in the goods market imply clearing in the money market account. In other words, the two clearing conditions in \eqref{clearingconds} are sufficient for clearing the money market account too.

The next theorem gives existence of a Radner equilibrium under the additional restriction $\theta^{(0)}_{2,0-}< \frac{g(0)}{D_0}$ where $g$ solves the below ODE \eqref{ODE_ineq}. This restriction is a generalization of the restriction in Eq. (6) in  \cite{BC1998}. In the following, we specialize  Theorem \ref{theorem_main_h} to the constant $A$ defined as
\begin{align}\label{Def_A}
A:= \frac{2 \beta+\sigma_D^2 -(1-\gamma) (2 \mu -\gamma  \sigma_D^2)}{\sigma_D^2}.
\end{align}
This particular expression for $A$ comes from equating the below ODE \eqref{ODE_ineq} with the ODE \eqref{KEY_ODE}. The restriction $A>1$ comes from \eqref{Dfinite} below. Instead of $h_{\xi_0}$, we use $h$ to denote the function produced by Theorem \ref{theorem_main_h} in the following. 

\begin{theorem}\label{ODE_g_existence}  Let $\gamma \in (0,1)$, $\sigma_D^2>0$, and $\beta >\frac{1}{2} (1-\gamma ) (2 \mu_D -\gamma  \sigma_D^2)$. 
\begin{enumerate}
\item There exists a nonincreasing and nonnegative solution $g \in\sC^1([0,1])\cap \sC^2([0,1))$ of the ODE
\begin{align}\label{ODE_ineq}
\begin{cases}
\beta g(y)= (1-\gamma)\mu_D g(y) - \tfrac12(1-\gamma)\gamma \sigma_D^2 g(y) +\mu_Y(y) g'(y) \\
\quad \quad \quad + \tfrac12\sigma_Y(y)^2g''(y)  + (1-\gamma)\sigma_D  \sigma_Y(y)g'(y)+(1-y)^{1-\gamma},\quad y\in(0,1),\\
g'(0) = 0,\quad g(1) =0.
\end{cases}
\end{align}

\item For $\theta^{(0)}_{2,0-}\in \big(0,\frac{g(0)}{D_0}\big)$, let $Y_0 \in (0,1)$ solve $g(Y_0) D_0(1-Y_0)^\gamma=\theta^{(0)}_{2,0-}$. Then, there exists a Radner equilibrium in which 
the  equilibrium interest rate process is $r_t=r(Y_t)\in \sL_{\text{loc}}^1$ and the equilibrium market price of risk process is $\kappa_t = \kappa(Y_t)\in \sL_{\text{loc}}^2$ for the deterministic functions
\begin{align}\label{randkappa}
\begin{split}
r(y) & := \beta +\gamma  \mu_D -\frac{1}{2} \gamma (\gamma +1) \sigma_D^2-\frac{\gamma  (\gamma +1) \sigma_D^2(1-y)}{2  y h(y)^2},\quad y\in(0,1),\\
\kappa(y) &:= \gamma  \sigma_D  \left(\frac{1-y}{y h(y)}+1\right),\quad y\in(0,1),
\end{split}
\end{align}
and the equilibrium consumption-rate processes in \eqref{primalval1} and \eqref{primalval2} are
\begin{align}\label{inhome15}
\hat{c}_{1,t}:=D_tY_t,\quad\hat{c}_{2,t}:=D_t(1-Y_t),\quad t\ge0.
\end{align}

\end{enumerate}
\end{theorem}

\proof[Proof of Theorem \ref{ODE_g_existence}(1)] Let $h$ be the solution produced by Theorem \ref{theorem_main_h}  for $A$ defined in \eqref{Def_A} and insert
\begin{align}\label{original_g}
g(y) := \frac{2}{\xi_0}e^{-\int_0^y \frac{h(q)}{1-q}dq} (1-y)^{-\gamma},\quad y \in [0,1),
\end{align}
and insert  $\sigma_Y$ and $\mu_Y$ from \eqref{mgmethod1bbq} into the ODE \eqref{ODE_ineq} to recover the ODE \eqref{KEY_ODE}. 

Next, we verify the boundary conditions in \eqref{ODE_ineq}. The representation \eqref{original_g} means that the boundary condition $g(1) = 0$ in \eqref{ODE_ineq}  requires
\begin{align}\label{need1}
\lim_{y\uparrow 1} \xi_0 e^{\int_0^y \frac{h(q)}{1-q}dq} (1-y)^{\gamma}=\infty,
\end{align}
Because $h(1) =1>\gamma$, Eq. \eqref{need1} holds. To see this,  we pick $\epsilon>0$ and $y_\epsilon \in (0,1)$ such that 
$$
\forall y\in (y_\epsilon,1): h(y) > \gamma+\epsilon.
$$
For $y\in(y_\epsilon,1)$, we have
\begin{align*}
e^{\int_0^y \frac{h(q)}{1-q}dq}(1-y)^\gamma &\ge e^{\int_{y_\epsilon}^y \frac{h(q)}{1-q}dq}(1-y)^\gamma \\
&\ge  e^{\int_{y_\epsilon}^y \frac{\gamma+\epsilon}{1-q}dq} (1-y)^\gamma\\
& = \frac{(1-y_\epsilon)^{\gamma+\epsilon}}{(1-y)^\epsilon}.
\end{align*}
Passing $y\uparrow 1$ gives the boundary condition $g(1)=0$. Because $g$ defined in \eqref{original_g} has derivative
\begin{align}\label{gprime}
g'(y)=\frac{2}{\xi_0}e^{-\int_0^y \frac{h(q)}{1-q}dq} \frac{1}{(1-y)^{1+\gamma}} \big( \gamma-h(y)\big),\quad y\in(0,1),
\end{align}
we see that the boundary condition $g'(0) = 0$ in \eqref{ODE_ineq} holds because $h(0) = \gamma$. 

Finally, because $h(y) \ge \gamma$ for $y\in [0,1]$, we see from \eqref{gprime} that $g'(y) \le 0$ for  $y\in (0,1)$.

The proof of Theorem \ref{ODE_g_existence}(2) is given in the next section. 

$\endproof$

\section{Proof of Theorem \ref{ODE_g_existence}(2)}
The duality techniques we use in the following proofs are standard and have been used in various contexts, see, e.g., the textbook \cite{KS1998}. Because the optimization problems \eqref{primalval} are on an infinite time horizon $t\in [0,\infty)$, there is a nontrivial tranversality condition as $t\to\infty$; see, e.g., Section 9D in \cite{Duffie2001}. As we shall see, the  tranversality  condition  is ensured by the parameter restriction $A>1$ for $A$ in \eqref{Def_A}, which is equivalent to $\beta >\frac12(1-\gamma ) \left(2 \mu_D -\gamma  \sigma_D^2\right)$. Among other things,  the restriction $A>1$ ensures that
\begin{align}\label{Dfinite}
\E\Big[ \int_0^\infty e^{-\beta t } D_t^{1-\gamma} dt \Big] =\frac{D_0^{1-\gamma} }{\beta -\frac12(1-\gamma) \left(2 \mu_D -\gamma \sigma_D^2\right)}<\infty,
\end{align}
where the geometric Brownian motion $D$ is from \eqref{dD}.

\subsection{State-price densities}
For an arbitrary process $\eta\in \sL^2_\text{loc}$, the process 
\begin{align}\label{dZeta}
dZ_{\eta,t} := -dZ_{\eta,t}\big( r(Y_t) dt + \eta_t dB_t\big),\quad Z_{\eta,0}\in (0,\infty),
\end{align}
is called a state-price density. Such density processes underly the duality theory we use to prove Theorem \ref{ODE_g_existence}(2) below. Any state-price density $Z_{\eta} $ has the defining property that 
the process $Z_{\eta,t} X_{i,t} + \int_0^t Z_{\eta,u} c_{i,u}du$, $t\ge0$, is a supermartingale for any admissible consumption process $c_i$ where $i=1$ or $i=2$ and $X_i$ is the corresponding wealth process in \eqref{dW1} or \eqref{dW2}.

In the below proofs, we establish the following first-order condition for investor 1
\begin{align}
\hat Z_{1,t} :=&\; e^{-\beta  t}( D_t Y_t)^{-\gamma}=e^{-\beta  t}\hat{c}_{1,t}^{-\gamma},\quad t\ge0,\label{inv001}\\
d\hat Z_{1,t}  =& - \hat Z_{1,t}\big( r(Y_t)dt + \kappa(Y_t) dB_t\big),\label{state-price1}
\end{align}
where we inserted $\hat c_{1} :=  DY$ from \eqref{inhome15}. The dynamics in \eqref{state-price1} follow from It\^o's lemma and 
\begin{align}\label{dDY1}
\begin{split}
d\hat c_{1,t} & = d D_tY_t \\
&= D_t\big(\mu_Y(Y_t)+\mu_D Y_t+\sigma_D\sigma_Y(Y_t) \big) dt + D_t \big(\sigma_Y(Y_t)+ \sigma_D Y_t \big)dB_t.
\end{split}
\end{align}
Similarly, we establish the following first-order condition for investor 2
\begin{align}
\hat Z_{2,t}  :=&\; e^{-\beta  t}\big( D_t (1-Y_t)\big)^{-\gamma}=e^{-\beta  t}\hat{c}_{2,t}^{-\gamma},\quad t\ge0,\label{inv002}\\
d\hat Z_{2,t}  =& - \hat Z_{2,t}\bigg( r(Y_t)dt -\gamma \sigma_D\frac{1-h(y)}{h(y)} dB_t\bigg),\label{state-price2} 
\end{align}
where we inserted  $\hat c_{2} := D(1-Y)$  from \eqref{inhome15}. The dynamics in \eqref{state-price2} follow from It\^o's lemma and 
\begin{align}\label{dDY}
\begin{split}
d \hat c_{2,t} & = d D_t(1-Y_t) \\
&=D_t\big(\mu_D-\mu_Y(Y_t) -\mu_D Y_t-\sigma_D\sigma_Y(Y_t)\big) dt + D_t\big( \sigma_D- \sigma_Y(Y_t) -\sigma_D Y_t \big) dB_t.
\end{split}
\end{align}
While the $dt$ terms of both $\frac{d\hat Z_1}{\hat Z_1}$ and $\frac{d\hat Z_2}{\hat Z_2}$ are identically equal to $-r(Y)$, the $dB$ terms differ. This is because investor 2 cannot access the stock market, which renders the model incomplete. 

\subsection{Investor 2's optimization problem}
Let $g$ denote the function from Theorem \ref{ODE_g_existence}(1) and  define  $r_t:= r(Y_t)$ and $\kappa_t := \kappa(Y_t)$ using the funtions in \eqref{randkappa}.
\begin{lemma}\label{lemma1} Let $\gamma \in (0,1)$, $\sigma_D^2>0$, and $\beta >\frac{1}{2} (1-\gamma ) (2 \mu_D -\gamma  \sigma_D^2)$.  
\begin{enumerate}
\item For $Y_0 \in (0,1)$, we have
\begin{align}\label{lem1}
 \E_t\left[\int_t^\infty e^{-\beta u} \big( D_u(1-Y_u)\big)^{1-\gamma} du\right] = e^{-\beta t}g(Y_t) D_t^{1-\gamma},\quad t\ge0.
\end{align}

\item When $Y_0 \in (0,1)$ satisfies $g(Y_0) D_0(1-Y_0)^\gamma=\theta^{(0)}_{2,0-}$, investor 2's value function is 
\begin{align}
\begin{split}
\sup_{c_2\in \sA_2} \tfrac1{1-\gamma}\E\left[\int_0^\infty e^{-\beta u} c^{1-\gamma}_{2,u} du\right] &=  \tfrac1{1-\gamma}\E\left[\int_0^\infty e^{-\beta u} \big( D_u(1-Y_u)\big)^{1-\gamma} du\right] \\
&= \tfrac1{1-\gamma}g(Y_0) D_0^{1-\gamma}.
\end{split}
\end{align}
\end{enumerate}

\end{lemma}
\proof  1: For $Y_0\in (0,1)$ and $n\in\N$ such that $Y_0 \in (\frac1n,1-\frac1n)$, we define the nondecreasing sequence of stopping times $\tau_n(\omega) := \inf\big\{s \ge t : Y_s(\omega)\not \in (\frac1n,1-\frac1n)\big\}$, $t\ge0$ and $\omega\in\Omega$, where $\inf\emptyset := \infty$. Lemma \ref{lemma1} and Lemma \ref{lem_vol} ensure that both $y=0$ and $y=1$ are inaccessible  and so $\P\big(Y_t \in (0,1) \;\forall t\ge0\big) = 1$. We replace $\tau_n$ with $\tau_n \land n$, to get that $\tau_n$ is bounded by $n$ uniformly in $\omega\in \Omega$, while  $\lim_{n\to\infty} \tau_n(\omega) = \infty$ continues to hold.

The ODE \eqref{ODE_ineq} and It\^o's lemma produce the following dynamics 
\begin{align}\label{FK1}
\begin{split}
&d\Big( e^{-\beta s}g(Y_s) D_s^{1-\gamma} + \int_0^s e^{-\beta u} \big( D_u(1-Y_u)\big)^{1-\gamma} du\Big)\\
&=e^{-\beta s}D_s^{1-\gamma}\Big\{\Big( (1-\gamma)\sigma_D g(Y_s)+  \sigma_Y(Y_s) g'(Y_s) \Big)dB_s\\
&+ \Big(-\beta g(Y_s) + (1-\gamma)\mu_D g(Y_s) - \tfrac12(1-\gamma)\gamma \sigma_D^2 g(Y_s) +\mu_Y(Y_s) g'(Y_s)\\
&+ \tfrac12\sigma_Y(Y_s)^2g''(Y_s)   + (1-\gamma)\sigma_D  \sigma_Y(Y_s) g'(Y_s) +(1-Y_s)^{1-\gamma}\Big) ds \Big\},\\
&=e^{-\beta s}D_s^{1-\gamma}\Big( (1-\gamma)\sigma_D g(Y_s)+  \sigma_Y(Y_s) g'(Y_s) \Big)dB_s,
\end{split}
\end{align}
for $s \in [0,\tau_n]$. We integrate \eqref{FK1} over $u\in[t,\tau_n]$ to get the representation
\begin{align}\label{FK11}
\begin{split}
&e^{-\beta \tau_n}g(Y_{\tau_n}) D_{\tau_n}^{1-\gamma} + \int_t^{\tau_n} e^{-\beta u} \big( D_u(1-Y_u)\big)^{1-\gamma} du \\
&= e^{-\beta t} g(Y_t) D_t^{1-\gamma}+ \int_t^{\tau_n} e^{-\beta u}D_u^{1-\gamma}\Big(  (1-\gamma)\sigma_D g(Y_u) +  \sigma_Y(Y_u)g'(Y_u) \Big)dB_u.
\end{split}
\end{align}
Taking conditional expectation and using the martingale property of the stopped $dB$ integral give 
\begin{align*}
&\E_t[e^{-\beta \tau_n}g(Y_{\tau_n}) D_{\tau_n}^{1-\gamma}] + \E_t\left[\int_t^{\tau_n} e^{-\beta u} \big( D_u(1-Y_u)\big)^{1-\gamma} du\right] = e^{-\beta t}g(Y_t) D_t^{1-\gamma}.
\end{align*}
The Monotone Convergence Theorem gives
\begin{align}\label{mon1}
\begin{split}
\lim_{n\to\infty} \E_t[e^{-\beta \tau_n}g(Y_{\tau_n}) D_{\tau_n}^{1-\gamma}] &+ \E_t\left[\int_t^\infty e^{-\beta u} \big( D_u(1-Y_u)\big)^{1-\gamma} du\right] \\
&= e^{-\beta t}g(Y_t) D_t^{1-\gamma}.
\end{split}
\end{align}
Because $D$ is a geometric Brownian motion, we have
\begin{align*}
e^{-\beta t} D_t^{1-\gamma} &= D_0^{1-\gamma}e^{-\beta t+(1-\gamma)(\mu_D - \frac12 \sigma_D^2)t + (1-\gamma)\sigma_D B_t}\\
&= D_0^{1-\gamma}e^{\big(-\beta +(1-\gamma)(\mu_D - \frac12 \sigma_D^2+\sigma_D \frac{B_t}t)\big)t  }.
\end{align*}
The law of the iterated logarithm gives $\lim_{t\to\infty}  \frac{B_t}t =0$. Furthermore, because $
\beta>(1-\gamma)(\mu_D - \frac12 \sigma_D^2)$, we have $\P$-a.s. the limit  $\lim_{t\to\infty} e^{-\beta t} D_t^{1-\gamma} =0$. Therefore, because $g$ is uniformly bounded, it suffices to show that the family of random variables $(e^{-\beta \tau_n}D_{\tau_n}^{1-\gamma})_{n\in \N}$ is uniformly integrable so that we can interchange limit and expectation in \eqref{mon1} to see $\P$-a.s. the following transversality condition
$$
\lim_{n\to\infty} \E_t[e^{-\beta \tau_n}g(Y_{\tau_n}) D_{\tau_n}^{1-\gamma}]=\E_t[\lim_{n\to\infty} e^{-\beta \tau_n}g(Y_{\tau_n}) D_{\tau_n}^{1-\gamma}]=0.
$$

To establish uniform integrability, we use the inequality  $\beta >(1-\gamma ) \left(\mu_D -\frac12\gamma  \sigma_D^2\right)$ to find $\epsilon>0$ but so small such that 
\begin{align}\label{epsi1}
\beta>(1-\gamma)\Big(\mu_D - \frac12 \sigma_D^2+ \frac12(1+\epsilon)(1-\gamma)\sigma_D^2\Big).
\end{align}
We can write
\begin{align*}
&(e^{-\beta t} D_t^{1-\gamma})^{1+\epsilon}\\
 &= D_0^{(1+\epsilon)(1-\gamma)}e^{- (1+\epsilon)\beta t+ (1+\epsilon)(1-\gamma)(\mu_D - \frac12 \sigma_D^2+\frac12(1+\epsilon)(1-\gamma)\sigma_D^2)t + (1+\epsilon)(1-\gamma)\sigma_D B_t- \frac12(1+\epsilon)^2(1-\gamma)^2\sigma_D^2 t}.
\end{align*}
The bound in \eqref{epsi1} and Cauchy-Schwartz give the inequality
\begin{align*}
&\E[(e^{-\beta \tau_n} D_{\tau_n}^{1-\gamma})^{1+\epsilon}] \\
&=D_0^{(1+\epsilon)(1-\gamma)}\E[e^{- (1+\epsilon)\big(\beta - (1-\gamma)(\mu_D - \frac12 \sigma_D^2+ \frac12(1+\epsilon)(1-\gamma)\sigma_D^2)\big)\tau_n + (1+\epsilon)(1-\gamma)\sigma_D B_{\tau_n}- \frac12(1+\epsilon)^2(1-\gamma)^2\sigma_D^2 \tau_n}]\\
&\le D_0^{(1+\epsilon)(1-\gamma)}\E[e^{ (1+\epsilon)(1-\gamma)\sigma_D B_{\tau_n}- \frac12(1+\epsilon)^2(1-\gamma)^2\sigma_D^2 \tau_n}]\\
&=D_0^{(1+\epsilon)(1-\gamma)},
\end{align*}
where the last equality uses Doob's optional sampling theorem (recall that $\tau_n$ is a bounded stopping time and that $e^{ p B_t- \frac12p^2 t}$ is a martingale for any $p\in \R$). De
la Valle\'e Poussin criterion gives uniform integrability.
\ \\

\noindent 2:  As in  \eqref{inhome15}, we define  $\hat{c}_{2} := D(1-Y)$ and we claim that the corresponding wealth processes is given by
\begin{align}\label{mgmethod1W}
\begin{split}
 \hat{X}_{2,t}&:=\frac{1}{\hat Z_{2,t}}\E_t \Big[\int_t^\infty e^{-\beta  u}\big(D_u(1-Y_u)\big)^{1-\gamma} du\Big] = D_tg(Y_t) (1-Y_t)^\gamma,\quad t\ge0,
\end{split}
\end{align}
where $\hat Z_{2}$ is defined in \eqref{inv002}. We need to show
\begin{align}\label{mgmethod1b}
\begin{split}
d \hat{X}_{2,t} &= \Big( \hat{X}_{2,t}r(Y_t) - \hat c_{2,t} \Big)dt,\quad \hat X_{2,0}= \theta^{(0)}_{2,0-}.
\end{split}
\end{align}
The initial condition $\hat X_{2,0} = \theta^{(0)}_{2,0-}$ holds by the assumption that $Y_0$ satisfies $g(Y_0) D_0(1-Y_0)^\gamma=\theta^{(0)}_{2,0-}$. By inserting 
$\hat{X}_{2}= Dg(Y) (1-Y)^\gamma$ from \eqref{mgmethod1W} and $\hat{c}_{2} = D(1-Y)$ from  \eqref{inhome15} into the dynamics \eqref{mgmethod1b}, the dynamics \eqref{mgmethod1b} follow as soon as we show 
\begin{align}\label{mgmethod1c}
\begin{split}
d\Big(\underbrace{D_tg(Y_t) (1-Y_t)^\gamma}_{\hat X_{2,t}}\Big) = \Big(\underbrace{ D_tg(Y_t) (1-Y_t)^\gamma}_{\hat X_{2,t}} r(Y_t) - \underbrace{D_t(1-Y_t)}_{\hat c_{2,t}} \Big)dt.
\end{split}
\end{align}
 It\^o's lemma and the functions in \eqref{randkappa} produce the dynamics \eqref{mgmethod1c}.

To see that $\hat c_2$ is optimal, we use the following standard duality argument. Set $U(x) := \frac1{1-\gamma} x^{1-\gamma}$ and let $V(t,y)$ be the Fenchel conjugate 
$$
V(t,b) := \sup_{x>0}\big( e^{-\beta t} U(x)-xb\big) = \frac{\gamma}{1-\gamma} e^{-\frac{\beta}{\gamma} t}b^{\frac{\gamma-1}\gamma}>0, \quad b\ge0.
$$ 
In the below inequalities, we use $\P(Y_t \le 1)=1$ for $t\ge0$, so that 
\begin{align*}
V(t,\hat Z_{2,t}) = \frac{\gamma}{1-\gamma} e^{-\beta t}   \big(D_t (1-Y_t)\big)^{1-\gamma} \le \frac{\gamma}{1-\gamma} e^{-\beta t}   D_t^{1-\gamma},\quad \P\text{-a.s.},
\end{align*}
which is integrable over $(t,\omega)\in [0,\infty)\times \Omega$ by \eqref{Dfinite}. Let $T\in(0,\infty)$ and let $c\in \sA_2$ be arbitrary with corresponding wealth process $X_{2}\ge0$ given in \eqref{dW2}. Then, 
\begin{align*}
\E \left[\int_0^T e^{-\beta t} U(c_t)dt\right] &\le \E\left[ \int_0^T\Big( V(t,\hat Z_{2,t}\big)  + \hat Z_{2,t}c_t\Big)dt\right]\\
&\le\E\left[ \int_0^T \Big( V(t,\hat Z_{2,t}\big)  + \hat Z_{2,t}c_t\Big)dt+\hat Z_{2,T}X_{2,T}\right]\\
&\le \E \left[\int_0^T V(t,\hat Z_{2,t})dt \right] + \hat Z_{2,0} X_{2,0}\\
&= \E \left[\int_0^T V(t,\hat Z_{2,t})dt \right] + \hat Z_{2,0} \theta^{(0)}_{2,0-},
\end{align*}
where the first inequality uses Fenchel's inequality, the second inequality follows from $X_{2,T}\ge0$, and the last inequality is produced by the supermartingale property of the state-price density $\hat Z_2$. We pass $T\uparrow \infty$ using the Monotone Convergence Theorem to get 
\begin{align*}
\E \left[\int_0^\infty e^{-\beta t} U(c_t)dt \right]&\le  \E \left[\int_0^\infty V(t,\hat Z_{2,t})dt \right] +  \hat Z_{2,0} \theta^{(0)}_{2,0-}\\
&=\E \left[\int_0^\infty V(t,\hat Z_{2,t}) dt \right]  + \hat Z_{2,0} \hat X_{2,0}\\
&= \E \left[\int_0^\infty  \Big( V(t,\hat Z_{2,t})+\hat Z_{2,t}\hat{c}_{2,t}\Big) dt \right] \\
&=\E \left[\int_0^\infty e^{-\beta t} U(\hat c_{2,t})dt \right],
\end{align*}
where the second equality uses \eqref{mgmethod1W} and the third equality uses \eqref{inv002}.
$\endproof$

\subsection{Investor 1's optimization problem}

Let $g$ denoted the function from Theorem \ref{ODE_g_existence}(1) and define  $r_t:= r(Y_t)$ and $\kappa_t := \kappa(Y_t)$ using the funtions in \eqref{randkappa}.

\begin{lemma}\label{lem_vol}  Let $\gamma \in (0,1)$, $\sigma_D^2>0$, $\beta >\frac{1}{2} (1-\gamma ) (2 \mu_D -\gamma  \sigma_D^2)$, and assume $Y_0 \in (0,1)$ satisfies $g(Y_0) D_0(1-Y_0)^\gamma=\theta^{(0)}_{2,0-}$.
\begin{enumerate}
\item There exists a volatility process $\sigma_{S}\in \mathcal{L}^2_\text{loc}$ such that the nonnegative processes
\begin{align}\label{mgmethod1}
\begin{split}
 S_t :&=\hat X_{1,t}+ \hat X_{2,t},\quad t\ge0,\\
 \hat{X}_{1,t}:&=\frac{1}{\hat Z_{1,t}}\E_t \Big[\int_t^\infty e^{-\beta  u}(D_uY_u)^{1-\gamma} du\Big],\quad t\ge0,
\end{split}
\end{align}
satisfy \eqref{dS} and 
\begin{align}\label{dW1hat}
\begin{split}
d\hat X_{1,t} & = r(Y_t) \hat X_{1,t} dt + S_{t}  \sigma_{S,t}\big(\kappa(Y_t)dt+ dB_t\big) - D_tY_tdt,\\ 
\hat X_{1,0}&= \theta^{(0)}_{1,0-} +S_0\in\R.
\end{split}
\end{align}
\item Investor 1's optimal consumption rate process is $\hat c_{1,t}:=D_tY_t$, $t\ge0$.
\end{enumerate}

\end{lemma}

\proof  1. The  Martingale Representation Theorem produces an integrand $\Delta\in \mathcal{L}^2_\text{loc}$ such that the second  equation in  \eqref{mgmethod1}  can be rewritten as
\begin{align}\label{spd_Xhat1}
\begin{split}
\hat Z_{1,t} \hat{X}_{1,t} + \int_0^t \hat Z_{1,u} D_uY_udu 
&=\E_t \Big[\int_0^\infty e^{-\beta  u}(D_uY_u)^{1-\gamma} du\Big]\\
&=\E \Big[\int_0^\infty e^{-\beta  u}(D_uY_u)^{1-\gamma} du\Big] + \int_0^t \Delta_s dB_s\\
&=\hat Z_{1,0}\hat X_{1,0} + \int_0^t \Delta_s dB_s.
\end{split}
\end{align}
It\^o's lemma produces the following dynamics of $\hat X_{1}$ in \eqref{mgmethod1}
\begin{align}\label{dW1_A}
\begin{split}
d \hat{X}_{1,t} &= d\left(\frac{ \hat Z_{1,t} \hat{X}_{1,t} }{ \hat Z_{1,t} } \right)\\
&=\Big(  \hat{X}_{1,t}\big(r(Y_t)+\kappa(Y_t)^2\big) -D_tY_t+\frac{\kappa(Y_t)\Delta_t}{\hat Z_{1,t}}\Big)dt +\Big( \hat{X}_{1,t}\kappa(Y_t)+\frac{\Delta_t}{\hat Z_{1,t}} \Big) dB_t.
\end{split}
\end{align}
By defining the volatility process
\begin{align}\label{sigmaSformula}
\sigma_{S,t} := \frac{1}{S_t}\Big( \hat{X}_{1,t} \kappa(Y_t)  +\frac{\Delta_t}{\hat Z_{1,t}} \Big),\quad t\ge0,
\end{align}
we see that the dynamics \eqref{dW1_A} equal those in \eqref{dW1}. In other words, $d\hat X_1$ are wealth dynamics for investor 1 when using  $\theta_{1,t}:=1$ and $c_{1,t}:=D_tY_t$ for $t\ge0$.

The initial value in  \eqref{dW1} follows from the first formula in \eqref{mgmethod1} and initial clearing in the money market (i.e.,  $\theta_{1,0-}^{(0)}  + \theta^{(0)}_{2,0-}=0$) because
\begin{align}\label{mgmethod14}
\begin{split}
\theta_{1,0-}^{(0)} + S_0 &= \theta_{1,0-}^{(0)}  + \hat{X}_{1,0} + \hat X_{2,0}\\
&= \theta_{1,0-}^{(0)}  +\hat{X}_{1,0} + \theta^{(0)}_{2,0-}\\
&=\hat{X}_{1,0}.
\end{split}
\end{align}

To see that \eqref{sigmaSformula} ensures that the dynamics $dS$ of the first equation in  \eqref{mgmethod1} agree with \eqref{dS}, we use  \eqref{dW1} and $\hat c_{2} = D(1-Y)$ to see
\begin{align}\label{mgmethod1r}
\begin{split}
dS_t&=d  \hat{X}_{1,t} +d \hat{X}_{2,t}\\
&=  r(Y_t) \hat X_{1,t} dt + S_{t}  \sigma_{S,t}\big(\kappa(Y_t)dt+ dB_t\big) - D_tY_tdt + \big(r(Y_t)\hat{X}_{2,t} -\hat c_{2,t}\big)dt\\
&=  r_t S_{t} dt + S_{t}  \sigma_{S,t}\big(\kappa(Y_t)dt+ dB_t\big) - D_tdt.
\end{split}
\end{align}
\ \\
\noindent 2. To see $\hat c_1 := DY$ and $\hat \theta_1:=1$ are optimal for investor 1, we again use the following standard duality argument as we did in the proof of Lemma \ref{lemma1}. Let $T>0$ and let $(c,\theta)\in \sA_1$ be arbitrary with corresponding wealth process $X_{1}$ given in \eqref{dW1}. Then, 
\begin{align*}
\E \left[\int_0^T e^{-\beta t} U(c_t)dt\right] &\le \E\left[ \int_0^T\Big( V(t,\hat Z_{1,t}\big)  + \hat Z_{1,t}c_t\Big)dt\right]\\
&\le\E\left[ \int_0^T \Big( V(t,\hat Z_{1,t}\big)  + \hat Z_{1,t}c_t\Big)dt+\hat Z_{1,T}X_{1,T}\right]\\
&\le \E \left[\int_0^T V(t,\hat Z_{1,t} )dt \right] + \hat Z_{1,0} X_{1,0},
\end{align*}
where the first inequality uses Fenchel's inequality, the second inequality follows from $X_{1,T}\ge0$, and the last inequality is produced by the supermartingale property of the state-price density $\hat Z_1$. We pass $T\uparrow \infty$ using the Monotone Convergence Theorem to get 
\begin{align*}
\E \left[\int_0^\infty e^{-\beta t} U(c_t)dt \right]&\le  \E \left[\int_0^\infty V(t,\hat Z_{1,t}) dt \right]  + \hat Z_{1,0} \hat X_{1,0}\\
&= \E \left[\int_0^\infty  \Big( V(t,\hat Z_{1,t})+\hat Z_{1,t}\hat{c}_{1,t}\Big) dt \right] \\
&=\E \left[\int_0^\infty e^{-\beta t} U(\hat c_{1,t})dt \right],
\end{align*}
where the first equality uses \eqref{mgmethod1} and the second equality uses \eqref{inv001}.
$\endproof$

\subsection{Remaining proof}

\proof[Proof of Theorem \ref{ODE_g_existence}(2)] The regularity conditions $r(Y_t)\in \sL_{\text{loc}}^1$ and $ \kappa(Y_t)\in \sL_{\text{loc}}^2$ follow from the functions in \eqref{randkappa} and the fact that $y=0$ and $y=1$ are inaccessible by Lemma \ref{lem_tech1} and Lemma \ref{lem2}. Optimality of \eqref{inhome15} and $\hat \theta_1:= 1$  follows from Lemma \ref{lemma1} and Lemma \ref{lem_vol}. Obviously, these control processes satisfy the clearing conditions in \eqref{clearingconds}.
$\endproof$

\appendix
\section{The failure of the comparison principle at $y=1$}\label{appA}
This appendix reports on the failure of the comparison principle at $y=1$. Theorem \ref{ODE_local_existence} uses Schauder's fixed-point theorem to construct a local solution of \eqref{KEY_ODE} starting from $y=0$. Alternatively, by redefining the set $\sX =\sX(y_0) \subset  \sC\big([y_0,1]\big)$ as those  continuous functions $h:[y_0,1]\to \R$ for which
\begin{align}\label{Lip1}
\sup_{y\in[y_0,1)}\frac{|1-h(y)|}{1-y} \le\frac{1-\gamma^2}{A-1}, 
\end{align}
for a constant $y_0 \in [0,1)$, it is possible to use Schauder's fixed-point theorem to construct a local solution of \eqref{KEY_ODE}  starting from $y=1$. However, as the next lemma shows, the comparison principle, i.e., the conclusion of Theorem \ref{Thm:Comparison},  can fail when we compare ODEs starting at $y=1$. This is the reason why we construct local solutions starting 
for $y=0$ in Theorem \ref{ODE_local_existence}. 

In the next proposition, the function $f\in \sC^1([0,1))\cap \sC([0,1])$  denotes the unique solution produced by setting $a_3 := - \gamma$ in Theorem \ref{Thm:ODEbound} of the ODE 
\begin{align}\label{f_ODEcomparison}
\begin{cases}
 &f'(y) =\frac{1+\gamma}{y}\Big( \gamma - f(y)\Big) +  \gamma f(y) \frac{1 - f(y)}{1-y},\quad y\in(0,1),\\
& f(0)=\gamma,\quad f(1) =1.
\end{cases}
\end{align}

\begin{proposition}\label{lem_appendix} Let $\gamma \in (0,1)$, $A>1$, let $f$ solve \eqref{f_ODEcomparison}, and let $h_{\xi_0}\in \sC^1([0,1])$ be as in Lemma \ref{lemma:xi0}. Then, we have
\begin{enumerate}
\item $f\notin\sC^1([0,1])$.
\item  $\lim_{y\uparrow1}f'(y) =\infty$, $\int_0^1 |f'(q)|dq<\infty$, and $\int_0^1 \frac{1-f(q)}{1-q} dq<\infty$.
\item $f(y)$ fails the comparison principle with $h_{\xi_0}(y)$ for $y$ close to $1$.
\end{enumerate}

\end{proposition}

\proof 1. To see $f\notin\sC^1([0,1])$, we argue by contradiction. By using  L'Hopital's rule and $f(1) =1$ in \eqref{f_ODEcomparison}, we get
$$
f'(1) =(1+\gamma)(\gamma - 1) +  \gamma \lim_{y\uparrow 1} f(y) \frac{1 - f(y)}{1-y}=\gamma^2 - 1 +  \gamma f'(1).
$$
This equation has the unique negative solution $f'(1) =-(1+\gamma)$. Theorem \ref{Thm:ODEbound} ensures that $f\le 1$. However, because $f(1) =1$, $f'(1) <0$ gives the contradiction $f(y)>1$ for $y$ close to 1.

 \noindent 2. {\bf Step 1/3:} We start by proving $\liminf_{y\uparrow1} \frac{1-f(y)}{1-y} =\infty$. 
Because $f(1)=1$, the first term on the right-hand-side of \eqref{f_ODEcomparison} has limit
\begin{align}\label{f_ODE_30}
\lim_{y \uparrow 1}\frac{1+\gamma}{y}\Big( \gamma - f(y)\Big)= -(1-\gamma)(1+\gamma)<0.
\end{align}
Set $\kappa := (1-\gamma)(1+\gamma)>0$ and consider any $\epsilon$ such that
$$
0<\epsilon  <\kappa.
$$
The limit in \eqref{f_ODE_30} gives $y_\epsilon\in(0,1)$ such that the solution of the ODE \eqref{f_ODEcomparison} satisfies
\begin{align}\label{f_ODE_35}
\begin{split}
f'(y) &\le -\kappa+\epsilon + \gamma f(y) \frac{ 1- f(y)}{1-y},\\
& \le  -\kappa+\epsilon+ \gamma  \frac{ 1- f(y)}{1-y},\quad y\in (y_\epsilon,1),
\end{split}
\end{align}
where  the second inequality uses $f\le 1$. Because
\begin{align}\label{f_ODE_36}
\begin{split}
\frac{d}{d y}\bigg(\frac{ 1-f(y)}{(1-y)^{\gamma}} \bigg)&=\gamma \frac{1-f(y)}{(1-y)^{1+\gamma}}  -\frac{f'(y)}{(1-y)^\gamma}  \ge  \frac{\kappa -\epsilon}{(1-y)^{\gamma}},
\end{split}
\end{align}
we have for $y\in (y_\epsilon,1)$
\begin{align}\label{f_ODE_37}
\begin{split}
\frac{1-f(y)}{(1-y)^\gamma} -\frac{1-f(y_\epsilon)}{(1-y_\epsilon)^\gamma} & =  \int_{y_\epsilon}^y\frac{d}{d q}\bigg(\frac{1-f(q)}{(1-q)^\gamma } \bigg) dq\\
&\ge(\kappa -\epsilon)  \int_{y_\epsilon}^y \frac{1}{(1-q)^\gamma} dq\\
&=-\frac{\kappa-\epsilon}{1-\gamma} \Big( (1-y)^{1-\gamma }-  (1-y_\epsilon)^{1-\gamma }\Big).
\end{split}
\end{align}

To argue by contradiction, we assume there is an increasing sequence $(y_n)_{n\in\N}\subset (y_\epsilon,1)$ converging to 1 such that
$$
\lim_{n\to\infty} \frac{1-f(y_n)}{1-y_n} = \liminf_{y\uparrow1} \frac{1-f(y)}{1-y} \in [0,\infty).
$$
In that case, we have
$$
\lim_{n\to\infty} \frac{1-f(y_n)}{(1-y_n)^\gamma} =\lim_{n\to\infty} \frac{1-f(y_n)}{1-y_n} (1-y_n)^{1-\gamma} = 0.
$$
In \eqref{f_ODE_37}, we replace $y$ by $y_n$ and pass $n\uparrow \infty$ to see
$$
-\frac{1-f(y_\epsilon)}{(1-y_\epsilon)^\gamma} \ge \frac{\kappa-\epsilon}{1-\gamma}  (1-y_\epsilon)^{1-\gamma }.
$$
By multiplying by $-(1-y_\epsilon)^\gamma$, we get the contradiction
$$
0\le 1-f(y_\epsilon) \le \frac{\epsilon-\kappa}{1-\gamma}  (1-y_\epsilon)<0,
$$
where the first inequality holds by Theorem \ref{Thm:ODEbound}.

\noindent{\bf Step 2/3:}  To prove $\lim_{y\uparrow1} f'(y)=\infty$, it suffices to show $ \liminf_{y\uparrow1}f'(y) =\infty$. By taking liminf in the ODE \eqref{f_ODEcomparison} and using $f(1) =1$, we get
$$
\liminf_{y\uparrow 1} f'(y) = -(1-\gamma)(1+\gamma) + \gamma \liminf_{y\uparrow 1}\frac{1-f(y)}{1-y} =\infty.
$$

\noindent{\bf Step 3/3:}  Consider $y_\epsilon\in(0,1)$ such that $f'(y)>0$ for $y \in (y_\epsilon,1)$. Then, $f(1) =1$ and the Monotone Convergence Theorem give 
$$
\int_{y_\epsilon} ^1 f'(q) dq = \lim_{y\uparrow 1}\int_{y_\epsilon} ^y f'(q) dq = \lim_{y\uparrow 1}  f(y) - f(y_\epsilon) =1-f(y_\epsilon) <\infty.
$$
Because $f(1)= 1$, by increasing $y_\epsilon$ if necessary, we can assume $\frac{f(y)}{\gamma} \ge \frac\gamma 2$ for $y\in (y_\epsilon,1)$. The ODE \eqref{f_ODEcomparison} gives 
\begin{align*}
\frac{1+\gamma}{y}f(y)+ f'(y) &=\frac{1+\gamma}{y} \gamma +  \gamma f(y) \frac{1 - f(y)}{1-y}\\
&\ge \frac{\gamma}2 \frac{1 - f(y)}{1-y},\quad y\in (y_\epsilon,1).
\end{align*}
Because the left-hand-side is integrable over $y\in  (y_\epsilon,1)$, the nonnegative function $ \frac{1 - f(y)}{1-y}$ is also integrable over $ y \in (y_\epsilon,1)$. For $y \in (0,y_\epsilon)$, the bounds $\gamma \le f(y) \le 1$ give integrability of $ \frac{1 - f(y)}{1-y}$.

\noindent 3.  The ODE for $h_{\xi_0}$ in \eqref{wODE} and $h_{\xi_0}(q)\le 1$ for all $q\in[0,1]$ give the lower bound
\begin{align}\label{C1_00w}
\begin{split}
h_{\xi_0}'(y) &=  \frac{1+\gamma}y \big(\gamma - h_{\xi_0}(y) \big)  + h_{\xi_0}(y) \frac{\gamma + \Big((A-\gamma) e^{\int_y^1\frac{1-h_{\xi_0}(q)}{1-q}dq} -A\Big) h_{\xi_0}(y)}{1-y}\\
&\ge   \frac{1+\gamma}y \big(\gamma - h_{\xi_0}(y) \big)  + \gamma h_{\xi_0}(y) \frac{ 1- h_{\xi_0}(y)}{1-y},\quad y\in (0,1).
\end{split}
\end{align}
To argue by contradiction, we assume that the comparison principle holds for $y$ near 1. Based on  \eqref{f_ODEcomparison} and \eqref{C1_00w}, the property $f(1) = h_{\xi_0}(1)=1$ gives $h_{\xi_0}(y)\le f(y)$ for $ y\in [0,1]$ and so
$$
\frac{1-h_{\xi_0}(y)}{1-y}\ge \frac{1-f(y)}{1-y}\ge0,\quad y\in (0,1).
$$
However, Lemma \ref{lemma:xi0}(7) ensures that $h_{\xi_0}'(y)$ is continuous at $y=1$ with limit $\frac{1-\gamma^2}{A-1}$ and so we get the contradiction
$$
\infty >\frac{1-\gamma^2}{A-1}=h_{\xi_0}'(1) =  \lim_{y\uparrow 1} \frac{1-h_{\xi_0}(y)}{1-y}\ge \lim_{y\uparrow 1} \frac{1-f(y)}{1-y}=\infty.
$$

$\endproof$


\begin{thebibliography}{9}       

\bibitem{Anderson-Raimondo} R. M. Anderson,  R. C. Raimondo, Equilibrium in continuous-time financial markets: Endogenously dynamically complete markets, Econometrica 76 (2008), 841--907.

\bibitem{BC1998} S. Basak, D. Cuoco, An equilibrium model with restricted stock market participation, Review of Financial Studies 11 (1998), 309--341.

\bibitem {brezis-peletier-termam86} H. Brezis, L. A. Peletier, D. A. Terman, 
A very singular solution of the heat equation with absorption. Arch. Rational
Mech. Anal. 95 (1986), no. 3, 185--209.

\bibitem{Chab2015} G. Chabakauri, Asset pricing with heterogeneous preferences, beliefs, and portfolio constraints,  Journal of Monetary Economics 75  (2015), 21--34.

\bibitem{Cherny_Engelbert} A. S. Cherny, H. J. Engelbert, Singular stochastic differential equations, 
Springer, Lecture Notes in Mathematics (2004).

\bibitem{CH1994} D. Cuoco, H. He, Dynamic equilibrium in infinite-dimensional economies with incomplete financial markets, working paper (1994).

\bibitem{Duffie2001} D. Duffie, Dynamic asset pricing theory, 3rd Ed., Princeton University Press (2001).

\bibitem {garcia-manasevich-yarur-98} M. Garc\'{\i}a-Huidobro, 
R. Man\'{a}sevich, C. S. Yarur,  On positive singular solutions for
a class of nonhomogeneous $p$-Laplacian-like equations,  J. Differential
Equations 145 (1998), no. 1, 23--51.

\bibitem{Helland1996} I. Helland, One-dimensional diffusion processes and their boundaries, working paper (1996).

\bibitem{Hug2012} J. Hugonnier, Rational asset pricing bubbles and portfolio 
constraints, Journal of Economic Theory 147 (2012), 2260--2302. 

\bibitem {iagar-laurencot13} R. G. Iagar,  P. Lauren\c{c}ot, 
Existence and uniqueness of very singular solutions for a fast diffusion
equation with gradient absorption. J. Lond. Math. Soc. (2) 87 (2013), no. 2, 509--529.

\bibitem{KS1988} I. Karatzas, S. Shreve, Brownian motion and stochastic calculus, 2nd Ed., Springer  (1988).

\bibitem{KS1998} I. Karatzas, S. Shreve, Methods of mathematical finance, Springer (1998).

\bibitem {leoni96} G. Leoni,  A very singular solution for the porous
media equation $u_{t}=\Delta(u^{m})-u^{p}$ when $0<m<1$, J. Differential
Equations 132 (1996), no. 2, 353--376.

\bibitem {mawhin-book87} J. Mawhin, Probl\`{e}mes de Dirichlet variationnels
non lin\'{e}aires. S\'{e}minaire de Math\'{e}matiques Sup\'{e}rieures, 104.
Presses de l'Universit\'{e} de Montr\'{e}al, Montreal, QC, 1987.

\bibitem {ni-serrin86} W.-M. Ni, J. Serrin, Nonexistence theorems for
singular solutions of quasilinear partial differential equations. Comm. Pure
Appl. Math. 39 (1986), no. 3, 379--399.

\bibitem{Pri2013} R. Prieto, Dynamic equilibrium with heterogeneous agents and risk constraints, working paper (2013).

\bibitem {pucci-serrin91} P. Pucci, J. Serrin, Continuation and
limit properties for solutions of strongly nonlinear second order differential
equations. Asymptotic Anal. 4 (1991), no. 2, 97--160.

\bibitem {pucci-serrin-book07} P. Pucci, J.  Serrin, Maximum
principles for elliptic partial differential equations. Handbook of
differential equations: stationary partial differential equations. Vol. IV,
355--483, Handb. Differ. Equ., Elsevier/North-Holland, Amsterdam, 2007. 

\bibitem{Revuz_Yor} D. Revuz, M. Yor, Continuous martingales and Brownian motion, 3rd Ed., Springer (1999).

\bibitem{Weston} K. Weston, Existence of an equilibrium with limited participation, Finance \& Stochastics (2024),  to appear.


\end{thebibliography}
\end{document}